\documentclass[%
 reprint,
superscriptaddress,
 amsmath,amssymb,
 aps
prb,
longbibliography
]{revtex4-1}

\usepackage{graphicx}
\usepackage{dcolumn}
\usepackage{bm}
\usepackage{multirow}
\usepackage{hyperref}
\usepackage{amsmath}%
\allowdisplaybreaks
\usepackage{bbm}
\usepackage{notes2bib}
\usepackage{float}
\usepackage{xcolor}

\newcommand{\ket}[1]{|#1\rangle}
\newcommand{\bra}[1]{\langle#1|}
\newcommand{\ave}[1]{\langle#1\rangle}

\begin{document}

\preprint{APS/123-QED}

\title{Identifying $C_{n}$-symmetric higher-order Topology and Fractional Corner Charge Using Entanglement Spectra}

\author{Penghao Zhu}
\affiliation{%
 Department of Physics and Institute for Condensed Matter Theory,\\
 University of Illinois at Urbana-Champaign, Urbana, Illinois 61801, USA
}%
\author{Kieran Loehr}
\affiliation{%
 Department of Physics,\\
 Cornell University, Ithaca, NY 14853, USA
}%
\author{Taylor L. Hughes}
\affiliation{%
 Department of Physics and Institute for Condensed Matter Theory,\\
 University of Illinois at Urbana-Champaign, IL 61801, USA
}%

\date{\today}

\begin{abstract}
We study the entanglement spectrum (ES) of two-dimensional $C_{n}$-symmetric second-order topological insulators (TIs). We show that some characteristic higher-order topological observables, e.g., the filling anomaly and its associated fractional corner charge, can be determined from the ES of atomic and fragile TIs. By constructing the relationship between the configuration of Wannier orbitals and the number of protected in-gap states in the ES for different symmetric cuts in real space, we express the fractional corner charge in terms of the number of protected in-gap states of the ES. We show that our formula is robust in the presence of electron-electron interactions as long as the interactions preserve $C_{n}$ rotation symmetry and charge-conservation symmetry. Moreover, we discuss the possible signatures of higher-order topology in the many-body ES. Our methods allow the identification of some classes of higher-order topology without requiring the usage of nested Wilson loops or nested entanglement spectra.
\end{abstract}

\maketitle


\section{\label{sec:intro}Introduction}
The entanglement spectrum (ES) is a useful diagnostic for topological phases \cite{PhysRevLett.101.010504}. For non-interacting Hamiltonians the single-particle entanglement spectrum can be calculated from the eigenvalues ($\{\xi\}$) of the single-particle correlation function 
\begin{equation}
\label{eq: corr}
C_{ij}(A)=\ave{c_{i}^{\dag}c_{j}},
\end{equation}
where $A$ is a subsystem of the whole system and $i,j \in A$ \cite{peschel2003calculation,PhysRevLett.101.010504,PhysRevB.84.205136,PhysRevLett.104.130502,PhysRevLett.105.115501,alexandradinata2011trace,hughes2011inversion,fang2013entanglement}. In previous work, it has been shown that the ES can identify Chern insulators and time-reversal symmetric TIs through the manifestation of spectral flow, analogous to the energy spectra of their edge states \cite{PhysRevLett.105.115501,alexandradinata2011trace}. Furthermore, topological crystalline insulators (TCIs) that have $C_{n}$ rotation or inversion symmetries can exhibit symmetry protected in-gap states in the single-particle ES, even though they may not exhibit mid-gap boundary modes in their energy spectrum \cite{hughes2011inversion,fang2013entanglement}.

Recent developments call for a re-examination of the entanglement spectra of TCIs. First, the initiation of the topological quantum chemistry classification paradigm has increased our understanding of the real-space structure of topological insulators and (obstructed) atomic limit insulators \cite{bradlyn2017topological}. Second, the study of TIs has been extended to higher-order topological insulators (HOTIs) that have symmetry-protected features at boundaries with co-dimension greater than 1, e.g., fractional corner charges in two dimensions \cite{benalcazar2017quantized,benalcazar2018quantization,wieder2018axion,van2018higher,lee2019fractional,lee2019higher,PhysRevB.97.205135}, and protected hinge modes in three dimensions \cite{Langbehn2017Reflection,schindler2018higher,benalcazar2018prb,Schindlereaat0346,PhysRevX.9.011012,PhysRevB.97.205135}. Conventionally, if a HOTI has gapped boundaries with co-dimension less than $n$, and symmetry-protected features on the boundaries with co-dimension $n$, then we call it an $n$-th order TI. Finally,
Ref. \onlinecite{benalcazar2018quantization} introduced a property called the \emph{filling anomaly}, that generically captures the charge fractionalization properties at the boundaries and corners of $C_{n}$ symmetric TCIs. 

Just as for strong TIs, one may expect that these features of TCIs can also be diagnosed by the single-particle ES. In fact, 
Refs. \onlinecite{Schindlereaat0346} and \onlinecite{PhysRevB.98.035147} showed that certain HOTI properties, which can be evaluated using nested Wilson loops \cite{benalcazar2017quantized,benalcazar2018prb}, can also be diagnosed via entanglement (e.g., using a nested ES) or the topology (e.g., entanglement polarization) of the many-body ground state of the entanglement Hamiltonian $\mathcal{H}_{e}$ \footnote{$\mathcal{H}_{e}$ is defined through the reduced density matrix $\rho_{A}=\frac{1}{Z_{e}}e^{-\mathcal{H}_{e}}$, of which the many-body ground state is determined by the single-particle ES in non-interacting systems \cite{hughes2011inversion,alexandradinata2011trace,fang2013entanglement}.}. Inspired by these results, one may ask if all higher-order features of HOTIs can be diagnosed by the nested ES instead of the single-particle ES. In this paper we answer this question by directly showing a relationship between the single-particle ES and the corner-induced filling anomaly that characterizes the higher-order fractional corner charges in two-dimensional $C_{n}$-symmetric second-order TIs. To accomplish this task we first relate the possible symmetric configurations of electronic Wannier orbitals (defined in Sec. \ref{subsec:Wannier}) with the number of \emph{protected} in-gap states in the ES (the results of Ref. \onlinecite{fang2013entanglement} play the key role in this relationship), and then use these relationships to relate the corner-induced filling anomaly, and thus the fractional corner charge, to the number of \emph{protected} in-gap states in the ES. 

The rest of our paper is organized as follows. In Sec \ref{subsec:Wannier}, we give a brief review about the Wannier orbital picture, and in Sec \ref{subsec:ES}, by re-interpreting the results of Ref. \onlinecite{fang2013entanglement} in the Wannier orbital picture, we relate the number of \emph{protected} in-gap states in the ES to the configuration of Wannier orbitals. Next, in Sec \ref{subsec: faes}, we build the relationship between the corner-induced filling anomaly and the configuration of Wannier orbitals, and then express the filling anomaly in terms of the number of \emph{protected} in-gap states in the ES. In Sec \ref{subsec: int}, we discuss the robustness of the corner-induced filling anomaly and the formulas derived in Sec \ref{subsec: faes} in the presence of symmetric interactions. In this process, we relate the number of protected in-gap states to the quasi-degeneracy of the many-body entanglement ground state. This provides us a way to monitor the change of the number of protected in-gap states in an adiabatic deformation when symmetric interactions are present. Finally, in Sec \ref{sec: zcl}, we study a zero correlation length limit, in which the entire ES for symmetric cuts can be clearly explained by the configuration of Wannier orbitals. From the intuitive pictures and numerical results in this section, we can gain a better intuition about the relationships among the configuration of Wannier orbitals, the ES, and the filling anomaly.

\section{\label{sec:ESWannier}Entanglement spectrum from Wannier orbital picture}

\subsection{\label{subsec:Wannier}Wannier orbital picture}
\begin{figure}[h]
\centering
\includegraphics[width=1\columnwidth]{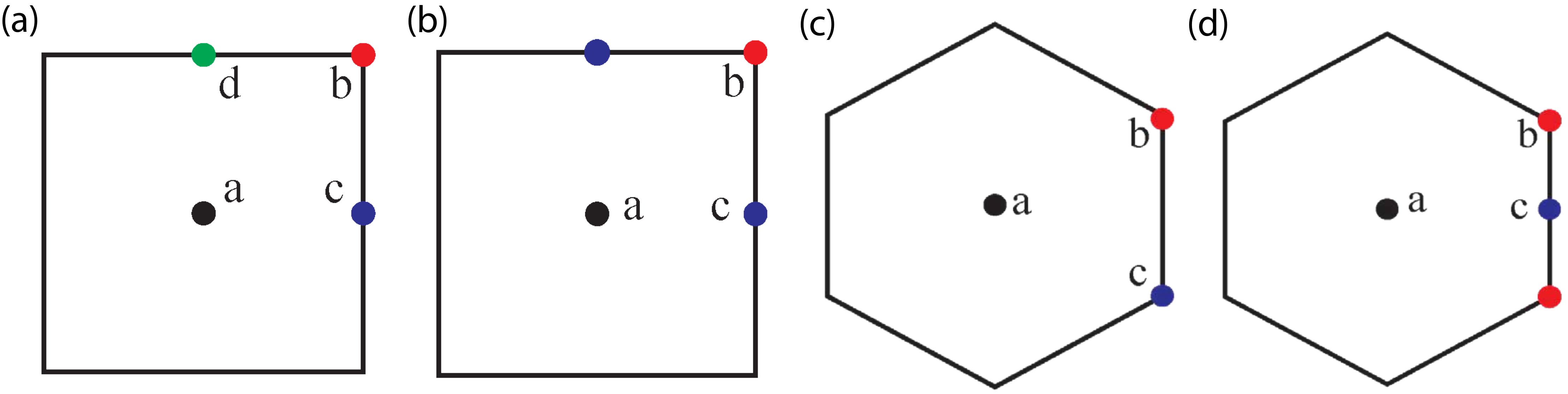}
\caption{Maximal Wyckoff positions in unit cells of  (a) $C_{2}$, (b) $C_{4}$, (c) $C_{3}$, and (d) $C_{6}$ symmetric two-dimensional Bravais lattices. Dots with the same color indicate that they are equivalent Wyckoff positions. We label different Wyckoff positions by letters $a$, $b$, $c$, and $d$.}
\label{fig:mwo}
\end{figure}

In order to discuss the relationship between the ES and the configuration of Wannier orbitals, it is important to introduce the Wannier orbital picture and notations we use at this point. If a set of bands can be represented in terms of
exponentially localized Wannier functions that preserve all the symmetries, we call them Wannier representable \cite{bradlyn2017topological,JenniferBuilding2018,po2017symmetry,po2018fragile,PhysRevX.7.041069}. In this terminology, \emph{atomic insulators} are defined to be TCIs with Wannier representable valence band(s). As such, from the valence band(s) of an atomic insulator, we can construct symmetric Wannier functions exponentially localized at high-symmetry points in each unit cell in real space. A high-symmetry point is defined to be a point ${\bf q}$, which is invariant under a subgroup $G_{\mathbf{q}}$ of the symmetry group $G$ of the Bravais lattice. We call the subgroup $G_{\mathbf{q}}$ the stabilizer group, and the high-symmetry point $\mathbf{q}$ the Wyckoff position.  Two Wyckoff positions $\mathbf{q}_{1}$ and $\mathbf{q}_{2}$ are said to be equivalent if their stabilizer groups are conjugate, \emph{i.e.}, there exists a $g\in G$ such that $G_{\mathbf{q}_{2}}=gG_{\mathbf{q}_{1}}g^{-1}$. If a Wyckoff position $\mathbf{q}$ has $M_{\mathbf{q}}$ equivalent Wyckoff positions (including itself), then we say it has multiplicity $M_{\mathbf{q}}$. If we cannot find a finite subgroup $H \subset G$ which satisfies $G_{\mathbf{q}} \subset H \subset G$, then we refer to $G_{\mathbf{q}}$ as a maximal stabilizer group, and correspondingly, we refer to $\mathbf{q}$ as a maximal Wyckoff position \cite{bradlyn2017topological,JenniferBuilding2018}. Since a set of symmetric Wannier orbitals localized at non-maximal Wyckoff positions can be continuously deformed to maximal Wyckoff positions while preserving the symmetry, we only study symmetric Wannier orbitals localized at maximal Wyckoff positions. Fig. \ref{fig:mwo} shows the maximal Wyckoff positions in unit cells of two-dimensional $C_{n}$-symmetric Bravais lattices for $n=2,3,4,6$. 

In two-dimensional $C_{n}$ symmetric lattices, the stabilizer group of a maximal Wyckoff position $\mathbf{q}$ is always isomorphic to a $C_{m_{\mathbf{q}}}$ rotation group with $m_{\mathbf{q}}\leqslant n$. Thus, we can define the angular momentum of Wannier orbitals localized at a maximal Wyckoff position $\mathbf{q}$ via the phase $e^{i(2\pi l+\pi F)/m_{\mathbf{q}}}$ gained upon a $C_{m_{\mathbf{q}}}$ rotation, where $l=0,1\ldots m_{\mathbf{q}}-1$, and $F=0,1$ corresponds to spinless and spin-$1/2$ cases, respectively. Because of translational symmetry, we only need to use the number of Wannier orbitals localized at different Wyckoff positions, having different angular momentum in each unit cell, $\{n^{l}_{\mathbf{q}}\}$, to describe an atomic insulator \cite{li2019fractional}. We will refer to $\{n^{l}_{\mathbf{q}}\}$ as the configuration of Wannier orbitals, and define $n_{\mathbf{q}}=\sum_{l}n^{l}_{\mathbf{q}}$ as the total number of Wannier orbitals localized at each Wyckoff position $\mathbf{q}$. Note that $n^{l}_{\mathbf{q}}$ of the Wyckoff position $\mathbf{q}$ with multiplicity $M_{\mathbf{q}}$ means the number of Wannier orbitals with angular momentum $l$ localized at \emph{one} of the $M_{\mathbf{q}}$ equivalent Wyckoff positions. For atomic insulators, it should be clear that we expect $n^{l}_{\mathbf{q}}\geqslant 0$. However, in the recently discovered fragile topological insulators, $n^{l}_{\mathbf{q}}$ can be generalized to take negative values, essentially implying that fragile topological insulators can be deformed into an atomic insulator by adding a set of Wannier representable bands to its valence bands \cite{po2018fragile,else2019fragile,bradlyn2019disconnected}.

\subsection{\label{subsec:ES}Relationship between the ES and the configuration of Wannier orbitals}
After reviewing the Wannier orbital picture we are ready to relate the properties of the ES to the configuration of Wannier orbitals.
To this end, we use the main conclusions in Ref. \onlinecite{fang2013entanglement} as follows.  In a $C_{n}$ symmetric lattice with $N$ unit cells, we denote a symmetric entanglement cut using $A^{m_{2}}_{1/m_{1}}$, where $m_{1}$ and $m_{2}$ are integer divisors of $n$, if it satisfies the three properties: (a) the number of unit cells in $A$ is $N/m_{1}$; (b) $A$ is a subset of the lattice with $C_{m_{2}}$ symmetry ; (c) the whole lattice can be generated by acting $C_{m_{1}m_{2}}$ on $A$. Ref. \onlinecite{fang2013entanglement} proves that for a cut $A^{m_{2}}_{1/m_{1}}$ in $C_{n}$-symmetric TCIs, the single-particle correlation operator is block diagonalized into $m_{2}$ blocks:
\begin{equation}
\label{eq:correlator}
C(A)=\frac{1}{m_{1}}\bigoplus_{r=0}^{m_{2}-1}\sum_{l=0}^{m_{1}-1}D_{l}^{r},
\end{equation}
where $(D_{l}^{r})^{2}=D_{l}^{r}$ is a projector that projects any state into the subspace in which states pick up the phase $e^{i(2\pi r+\pi F)/m_{2}}$ under $C_{m_{2}}$, and  the phase $e^{i(2\pi (r+lm_{2})+\pi F)/m_{1}m_{2}}$ under $C_{m_{1}m_{2}}$. Utilizing Eq. \eqref{eq:correlator}, it has been proved that there are $\sum_{r}\mathrm{max} _{l,l^{\prime}}|\mathrm{dim}(D^{r}_{l})-\mathrm{dim}(D^{r}_{l^{\prime}})|$ protected in-gap states in the range of $\left[1/m_{1},1-1/m_1\right]$ in the ES for the cut $A^{m_{2}}_{1/m_{1}}$ \cite{fang2013entanglement}. For a brief review of the proof, see Appendix \ref{app:review}. If we define $z_{i}$ as the number of occupied states with angular momentum $i=0,1,\ldots n-1$, then $(z_{0},z_{1},...,z_{n-1})$ forms a $\mathbb{Z}^{n}$ index. Ref. \onlinecite{fang2013entanglement} shows that $\mathrm{dim}(D^{r}_{l})$ can be expressed in terms of the $\mathbb{Z}^{n}$ index, and thus the number of protected in-gap states in the ES can be written in terms of the $\mathbb{Z}^{n}$ index. 

\begin{figure}[h]
\centering
\includegraphics[width=1\columnwidth]{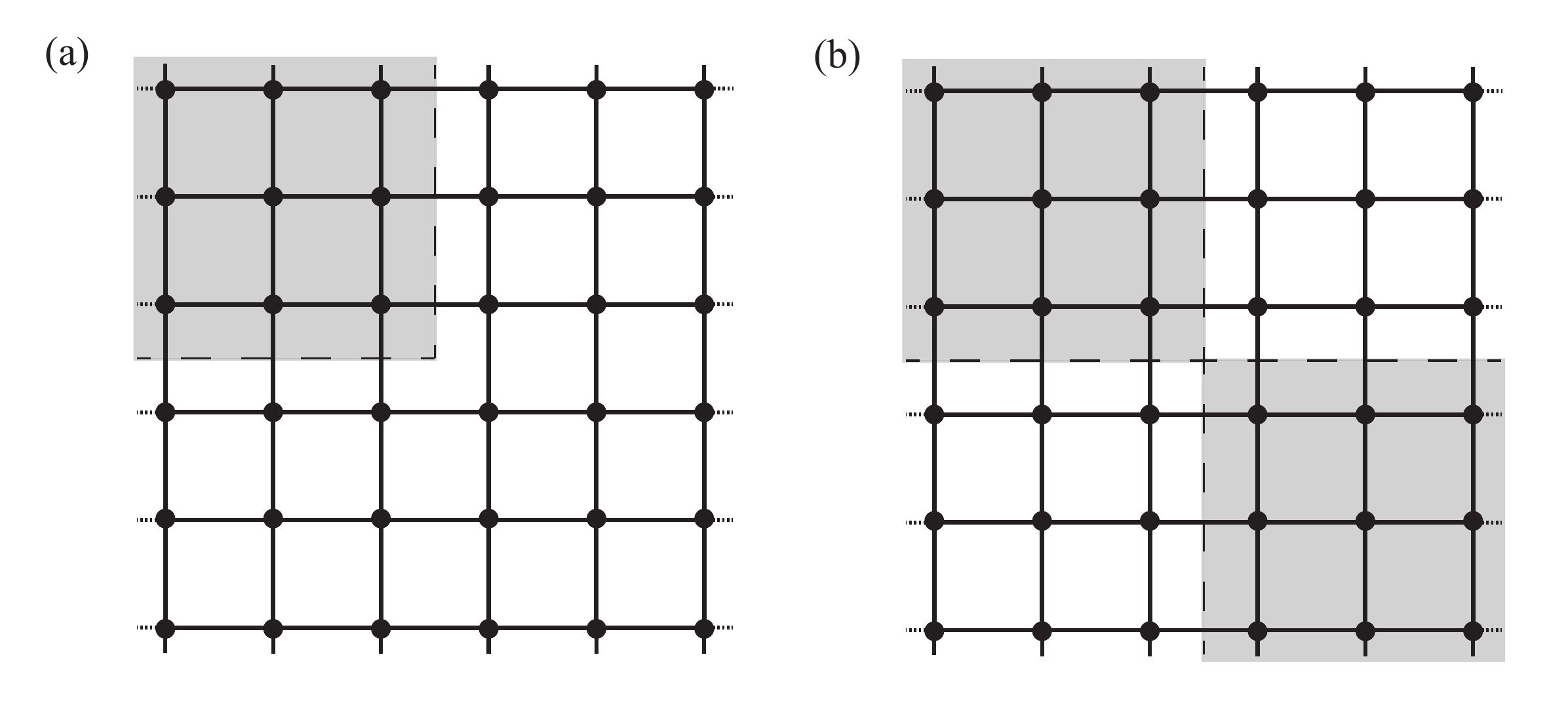}
\caption{(a) The cut $A^{1}_{1/4}$ and (b)$A^{2}_{1/2}$ in a $C_{4}$ symmetric square lattice with rotation center localized at Wyckoff position $b$.}
\label{fig:C4twoCuts}
\end{figure}

As examples, let us consider the cuts $A^{1}_{1/4}$ and $A^{2}_{1/2}$ in a  $C_{4}$ symmetric square lattice (illustrated in Fig. \ref{fig:C4twoCuts}). According to Table I in Ref \onlinecite{fang2013entanglement} (partially shown in Table \ref{tab:tab1ref8} in Appendix \ref{app:review}), we know that there would be $\mathrm{max}_{i,j=0,1,2,3}|z_{i}-z_{j}|$ protected in-gap states in the range $\left[1/4,3/4\right]$ in the ES for the cut $A^{1}_{1/4}$, and there would be $|z_{0}-z_{2}|+|z_{1}-z_{3}|$ protected in-gap states with eigenvalue $1/2$ in the ES of the cut $A^{2}_{1/2}$. 

Importantly, the $\mathbb{Z}^{n}$ index can be derived from the configuration of Wannier orbitals in atomic insulators. To illustrate this, first consider the case with a $C_4$ rotation center located at Wyckoff position $b$ in a $2L\times 2L$ ($L \rightarrow \infty$) square lattice (as shown in Fig. \ref{fig:C4twoCuts}). In this case, all Wannier orbitals localized at Wyckoff positions $a$, $b$ (other than the rotation center) and $c$ always appear in quadruplets which form a regular representation of the $C_{4}$ point group, \emph{i.e.}, the $\mathbb{Z}^{n}$ index coming from each quadruplet is ${z_{i}=1}$ for all $i=0,1,2,3$. Only Wannier orbitals localized at the rotation center (Wyckoff position $b$) contribute to any differences in the $z_{i}$, and thus contribute to the number of protected in-gap states in the ES. For example, in the ES for the cut $A^{1}_{1/4}$ shown in Fig. \ref{fig:C4twoCuts}(a), we have

\begin{equation}
\label{eq:cwops}
\begin{aligned}
\mathrm{no}.\left[1/4,3/4\right]_{p,4,1}^{(4)}&=\mathrm{max}_{i,j=0,1,2,3}|z_{i}-z_{j}|
\\
&=\mathrm{max}_{l,l^{\prime}=0,1,2,3}|n_{b}^{l}-n_{b}^{l^{\prime}}|,
\end{aligned}
\end{equation}
where $\mathrm{no}.\left[1/4,3/4\right]_{p,4,1}^{(4)}$ is the number of protected in-gap states in the ES for the cut $A^{1}_{1/4}$. The superscript on the bracket denotes four-fold rotation symmetry, and the subscript means that we are counting only \emph{protected} in-gap states in the ES for the cut $A^{1}_{1/4}$. Also, in the ES for the cut $A^{2}_{1/2}$ shown in Fig. \ref{fig:C4twoCuts}(b), we have 
\begin{equation}
\label{eq:onehalf}
\mathrm{no}.\frac{1}{2}_{p,2,2}^{(4)}=|z_{0}-z_{2}|+|z_{1}-z_{3}|=|n^{0}_{b}-n^{2}_{b}|+|n^{1}_{b}-n^{3}_{b}|,
\end{equation}
where $\mathrm{no}.\frac{1}{2}_{p,2,2}^{(4)}$ is the number of protected in-gap states with eigenvalue $1/2$ in the ES for the cut $A^{2}_{1/2}$.

In addition, according to Eq. \eqref{eq:correlator}, the reduced correlation operator for the cut $A^{1}_{1/4}$ ($m_{1}=4$, $m_{2}=1$) is 
\begin{equation}
C(A)=\frac{1}{4}(D_{0}+D_{1}+D_{2}+D_{3}),
\end{equation} 
where $D_{l}$ is a projector of the subspace in which states pick up a phase $e^{i(2\pi l+\pi F)/4}$ under $C_{4}$ rotation.
Note that when $m_{2}=1$ the superscript $r$ in Eq. \eqref{eq:correlator} can only be zero, and thus has been suppressed in the above equation. Since each $D_{l}$ satisfies $D_{l}^{2}=D_{l}$, its eigenvalues are either $0$ or $1$. If one of them has dimension greater than all of the others, then there are protected in-gap states with eigenvalue $1/4$, of which the number is
\begin{equation}
\label{eq:onefourth}
\begin{aligned}
\mathrm{no}.\frac{1}{4}_{p,4,1}^{(4)}&=\mathrm{dim}(D^{first})-\mathrm{dim}(D^{second})=z^{first}-z^{second}
\\
&=n_{b}^{first}-n_{b}^{second},
\end{aligned}
\end{equation}
where the superscripts $first$ and $second$ mean the highest and the second highest values of of $\mathrm{dim}(D_{i})$ (or $z_{i}$ and $n^{l}_{b}$) with $i,l=0,1,2,3$. 


Next, let us discuss the case with the rotation center localized at Wyckoff position $c$. In this case, we have only $C_{2}$ rotation symmetry, and all Wannier orbitals localized at Wyckoff positions $a$, $b,$ or $c$ (other than the rotation center) always come in pairs. Each of these pairs has a $\mathbb{Z}^{2}$ index ${z_{0}=1,z_{1}=1}$. Again, according to Table I in Ref. \onlinecite{fang2013entanglement}, there are $|z_{0}-z_{1}|$ protected $1/2$ states in the ES for the cut $A^{1}_{1/2}$. Thus, we have 
\begin{equation}
\label{eq:onehalfc}
\mathrm{no}.\frac{1}{2}_{p,2,1}^{(2)}=|z_{0}-z_{1}|=|n^{0}_{c}-n^{1}_{c}|,
\end{equation}
where $\mathrm{no}.\frac{1}{2}_{p,2,1}^{(2)}$ is the number of protected in-gap states with eigenvalues $1/2$ in the ES for the cut $A^{2}_{1/2}$ when there is two-fold rotation symmetry. For $C_{2}$, $C_{3}$, and $C_{6}$ symmetric atomic insulators, we can have similar relationships between the number of protected in-gap states in the ES and the configuration of Wannier orbitals $n_{\bf q}^{l}$ by choosing different symmetric cuts and rotation centers localized at the various maximal Wyckoff positions. Details are discussed in Appendix \ref{app:ingap}. 

As an example, in Fig. \ref{fig:atomic}, we numerically calculate the ES for the cut $A^{1}_{1/2}$ in a $10\times 10$ open square lattice for the $h^{(4)}_{2b}$ model ($n_{b}^{0}=n_{b}^{2}=0$, $n_{b}^{1}=n_{b}^{3}=1$, and $\mathrm{others}=0$ at half filling) introduced in Ref. \onlinecite{benalcazar2018quantization}:
\begin{equation}
\label{eq:h2b}
\begin{aligned}
&h^{(4)}_{2b}(\mathbf{k})=e^{i(k_{x}-k_{y})}\hat{c}^{\dag}_{3,\mathbf{k}}\hat{c}_{1,\mathbf{k}}+e^{i(k_{y}+k_{x})}\hat{c}^{\dag}_{2,\mathbf{k}}\hat{c}_{4,\mathbf{k}}
\\
&+\lambda_{1}\sum_{a=1}^{4}\hat{c}^{\dag}_{[a+1] ,\mathbf{k}}\hat{c}_{[a],\mathbf{k}}+\lambda_{2}(\hat{c}^{\dag}_{3,\mathbf{k}}\hat{c}_{1,\mathbf{k}}+\hat{c}^{\dag}_{2,\mathbf{k}}\hat{c}_{4,\mathbf{k}})
\\& \ \ \ \ \ +h.c.,
\end{aligned}
\end{equation}
where $c_{a,\mathbf{k}}$ is the Fourier transformation of $c_{a,\mathbf{R}}$, which is the annihilation operator of a spinless electron on sublattice $a$ in the unit cell located at $\mathbf{R}$ on a square lattice, and $[a]\equiv a\ \mathrm{mod} \ 4$. According to Appendix \ref{app:ingap}, we should find $|n_{b}^{0}+n_{b}^{2}-n_{b}^{1}-n_{b}^{3}|=2$ in-gap states with eigenvalue $1/2$, which is consistent with our numerical result.
\begin{figure}[h]
\centering
\includegraphics[width=1\columnwidth]{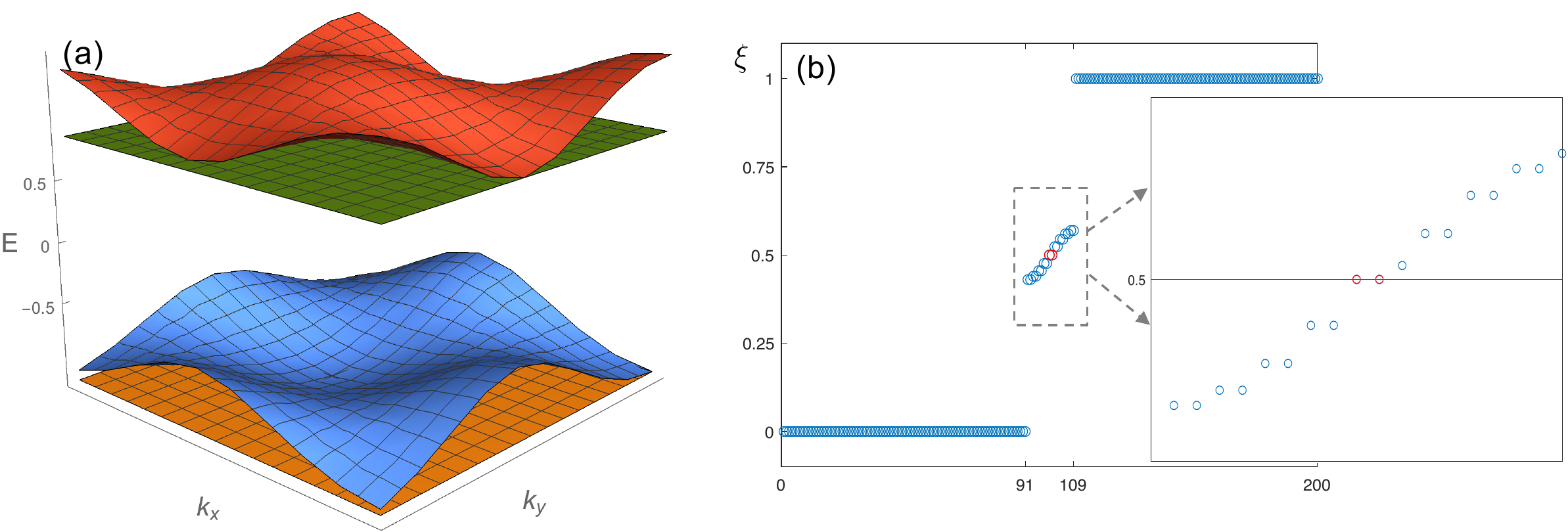}
\caption{(a) Energy spectrum of $h_{2b}^{(4)}$ for $\lambda_{1}=\lambda_{2}=0.15$. The lowest two bands are Wannier representable, and have the Wannier orbital configuration $n_{b}^{0}=n_{b}^{2}=0$, $n_{b}^{1}=n_{b}^{3}=1$, and $\mathrm{others}=0$. (b) The ES of $h_{2b}^{(4)}$ with $\lambda_{1}=\lambda_{2}=0.15$ for the cut $A^{1}_{1/2}$ in a $10\times 10$ open square lattice. The x-axis represents the sequence number of eigenstates of the reduced correlation operator, and the y-axis represents the corresponding eigenvalues. There are only two in-gap states with eigenvalue $1/2$ emphasized by two red circles. Thus, they must be the predicted two protected in-gap states with eigenvalue $1/2$.}
\label{fig:atomic}
\end{figure}

\begin{figure}[h]
\centering
\includegraphics[width=1\columnwidth]{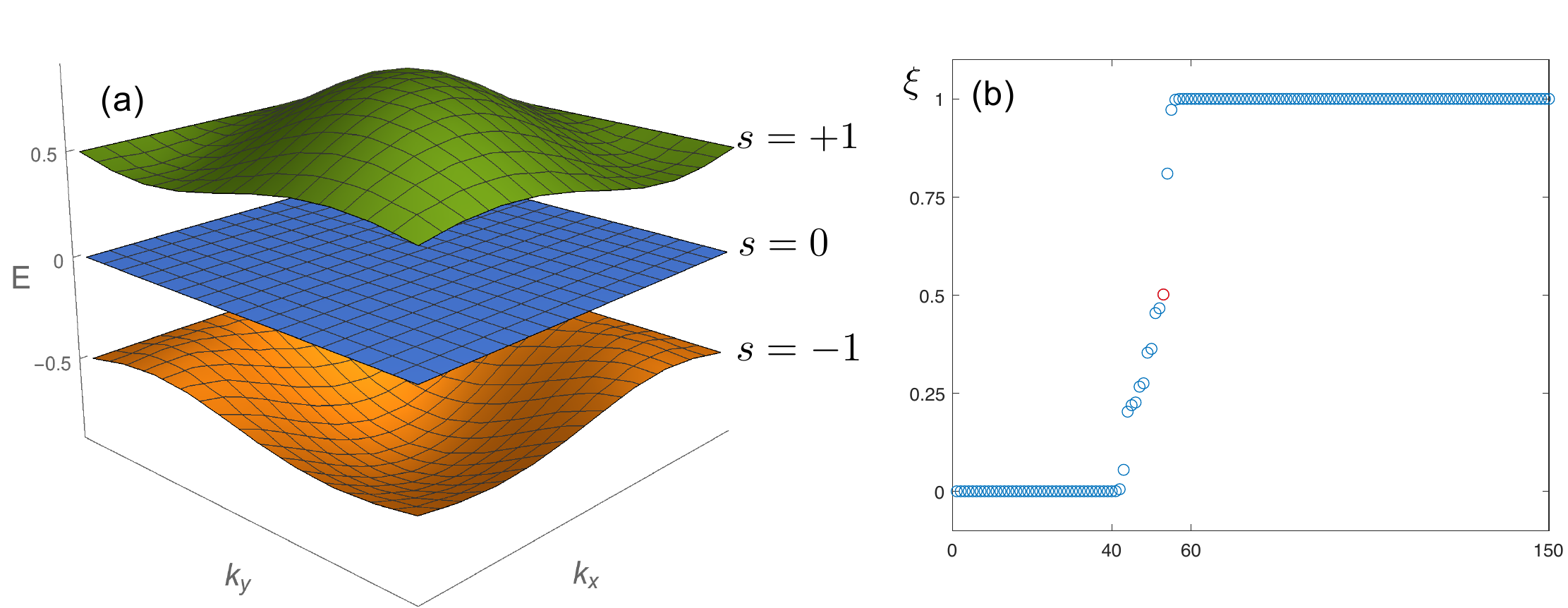}
\caption{(a) Energy spectrum of $H_{\mathbf{k}}^{(1)}$ for $t=i/4$; (b) the ES of the fragile TI described by Eq. \eqref{eq: fragile} with $t=i/4$ for the cut $A^{1}_{1/2}$ in a $10\times 10$ open square lattice. The x-axis represents the sequence number of eigenstates of the reduced correlation operator, and the y-axis represents the corresponding eigenvalues. There is only one in-gap states with eigenvalue $1/2$ emphasized by one red circle. Thus, it must be the predicted protected in-gap state with eigenvalue $1/2$.}
\label{fig:fragile}
\end{figure}

The relationships derived above are also applicable for fragile TIs that have negative $n_{\mathbf{q}}^{l}$'s in their configurations of Wannier orbitals. Let us take the $C_{2}$ symmetric tight-binding model in Ref. \onlinecite{else2019fragile} (shown in Eq.\eqref{eq: fragile} below) as an example to illustrate the idea. We first define
\begin{equation}
\label{eq: fragile1}
\begin{aligned} H_{\mathbf{k}}^{(1)}=& t\left(1+e^{i k_{x}}\right) \hat{c}_{2, \mathbf{k}}^{\dagger} \hat{c}_{1, \mathbf{k}}+t\left(1+e^{i k_{y}-i k_{x}}\right) \hat{c}_{3, \hat{k}}^{\dagger} \hat{c}_{2, \mathbf{k}} \\ &+t\left(1+e^{-i k_{y}}\right) \hat{c}_{1, \mathbf{k}}^{\dagger} \hat{c}_{3, \mathbf{k}}+\text { h.c. } ,\end{aligned}
\end{equation}
where $c_{a,\mathbf{k}}$ is the Fourier transformation of $c_{a,\mathbf{R}}$, which is the annihilation operator of a spinless electron on sublattice $a$ in the unit cell located at $\mathbf{R}$ on a square lattice. If we diagonalize $H_{\mathbf{k}}^{(1)}$, we can write
\begin{equation}
\label{eq: fragile2}
H_{\mathbf{k}}^{(1)}=\sum_{s=0, \pm 1} \epsilon_{s, \mathbf{k}} \hat{\gamma}_{s, \mathbf{k}}^{\dagger} \hat{\gamma}_{s, \mathbf{k}},
\end{equation}
where $s=0,\pm 1$ labels the three energy bands of $H_{\mathbf{k}}^{(1)}$. We plot the $\epsilon_{s, \mathbf{k}}$ for $t=i/4$ in Fig. \ref{fig:fragile}(a). According to Ref. \onlinecite{else2019fragile}, the pair of $s=\pm 1$ bands together form a set of fragile bands, for which the configuration of Wannier orbitals is $\{n_{a}^{0}= n^{0}_{c}=1, n_{b}^{0}=-1, \mathrm{others}=0\}$.
By adding $H_{\mathbf{k}}^{(2)}=-\sum_{a=1}^{3} \hat{c}_{a, \mathbf{k}}^{\dagger} \hat{c}_{a, \mathbf{k}}+2 \hat{\gamma}_{0, \mathbf{k}}^{\dagger} \hat{\gamma}_{0, \mathbf{k}}$ to $H_{\mathbf{k}}^{(1)}$, we can force the $s=\pm 1$ bands to lie below the $s=0$ band, and thus we can construct a fragile TI described by 
\begin{equation}
\label{eq: fragile}
H_{fragile}(\mathbf{k})=H_{\mathbf{k}}^{(1)}+H_{\mathbf{k}}^{(2)},
\end{equation} where we fill the lower two bands.

Let us now consider the ES of this fragile TI for the cut $A^{1}_{1/2}$ in a $2L\times 2L$ square lattice with open boundary conditions along both directions. According to Appendix \ref{app:ingap}, the number of protected in-gap states with eigenvalue $1/2$ should be $|n_{b}^{0}-n_{b}^{1}|$. Since the fragile TI described by Eq.\eqref{eq: fragile} has  $n_{b}^{0}=-1$ and $n_{b}^{1}=0$, we expect to have one one-half state in the ES. This is confirmed by our numerical calculations in a $10\times 10$ open square lattice as shown in Fig. \ref{fig:fragile}(b). Hence, the properties of the protected modes in the ES can be determined from the Wannier configurations for rotationally symmetric atomic and fragile insulators.

\section{Filling anomaly and entanglement spectrum}
\label{sec:fa}
\subsection{Corner-induced filling anomaly in terms of the number of protected in-gap states in the ES }
\label{subsec: faes}
\begin{figure}[htp]
\centering
\includegraphics[width=1\columnwidth]{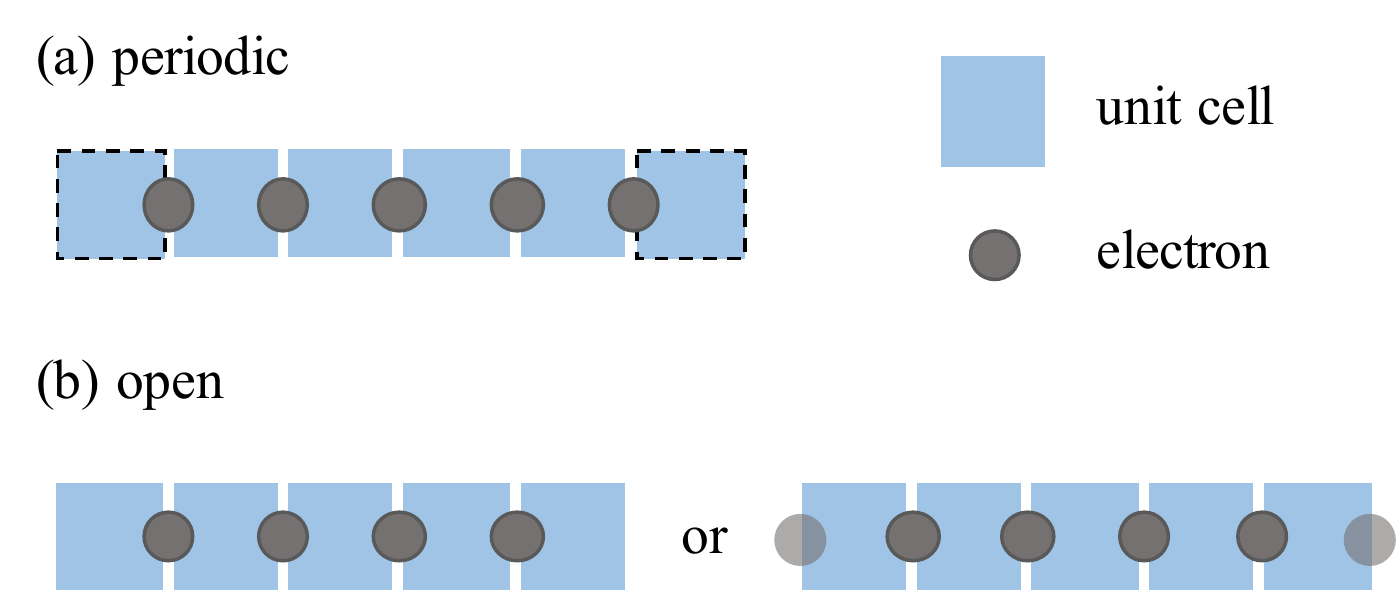}
\caption{In an inversion symmetric nontrivial SSH chain with $N$ unit cells, if we have (a)periodic boundary condition, $N$ Wannier orbitals are occupied, $i.e.$, $N_{occ,p}=N$, and the unit cells with dotted outline are identified; (b)open boundary condition, $N-1$ or $N+1$ Wannier orbitals are occupied, $i.e.$, $N_{occ,o}=N-1$ or $N+1$, and the dimmer circles represent edge states.}
\label{fig:fa}
\end{figure}

Knowing the relationship between the configuration of Wannier orbitals and the ES, we are now ready to discuss the filling anomaly. The filling anomaly (see Ref. \onlinecite{benalcazar2018quantization}) is a property of some $C_n$ invariant TCIs that directly determines the amount of fractional edge and/or corner charge in open sample geometries. One can define the filling anomaly $\eta$ as
\begin{equation}
\label{eq:definitionfa}
\eta=N_{occ,p}-N_{occ,o} \ \mathrm{mod} \ n,
\end{equation}
where $N_{occ,p}$ and $N_{occ,o}$ are the numbers of occupied states of the TCIs under periodic and open boundary conditions, respectively, under certain conditions. To illustrate the definition, we take the inversion symmetric (i.e., $C_2$ symmetric) nontrivial Su-Schrieffer-Heeger (SSH) chain \cite{PhysRevLett.42.1698} with $N$ unit cells for example. If we have periodic boundary conditions (as shown in Fig. \ref{fig:fa} (a)), then we can have a gapped, symmetric, neutral insulator if we fill $N_{occ,p}=N$ Wannier orbitals; for the obstructed phase these orbitals are located in between two unit cells. However, if we have open boundary conditions (as shown in Fig. \ref{fig:fa} (b)), we must fill $N_{occ,o}=N-1$ or $N+1$ Wannier orbitals to preserve the inversion/$C_2$ symmetry and the energy gap. This violates neutrality and leads to a filling anomaly of $\eta=1.$ If we take $\eta/n$ (for $C_n$ symmetric systems) we find the amount of fractional charge induced at edges or corners. For this case the filling anomaly is related to the ends of the chain and thus we have a fractional charge of $e/2\mod e$ on the ends.  

More generally in two dimensions, both edges and corners of atomic and fragile insulators can be associated to the filling anomaly depending on the underlying Wannier configuration. However, as discussed in Ref. \onlinecite{benalcazar2018quantization}, when the bulk polarization is zero, the filling anomaly defined in Eq. \eqref{eq:definitionfa} has contributions from only the corners, which we will call a corner-induced filling anomaly. A corner-induced filling anomaly of $\eta$ in a $C_n$ symmetric insulator implies a fractional charge of $\eta/n$ in each $2\pi/n$ sector of an open $C_n$ symmetric lattice. In simple geometries this fractional charge is localized on the corners of the sample \cite{benalcazar2018quantization}. Since we are primarily interested in these higher-order corner features, let us now constrain the polarization to vanish. The requirement that the bulk polarization vanishes places constraints on the allowed Wannier configurations: one must have $n_{b}$ and $n_{c}$ simultaneously odd or even in the $C_{4}$ symmetric case, $n_{b}\ \mathrm{mod}\ 2=n_{c}\ \mathrm{mod}\ 2=n_{d}\ \mathrm{mod}\ 2$ in the $C_{2}$ symmetric case, and $n_{b}\ \mathrm{mod}\ 3=n_{c}\ \mathrm{mod}\ 3$ in the $C_{3}$ symmetric case. We note that $C_6$ symmetry forces the bulk polarization to vanish in general so any symmetric Wannier configuration is allowed. This is because $C_{2}$ symmetry requires the polarization to be $0$ or $\frac{1}{2},$ and $C_{3}$ symmetry requires the polarization to be $0$, $\frac{1}{3}$, and $\frac{2}{3}$. Thus, in the $C_{6}$ symmetric case which contains both $C_{2}$ and $C_{3}$ symmetry, the only possible polarization is zero, and there are no further constraints on the configuration of Wannier orbitals. 

Now we can evaluate the corner-induced filling anomalies in terms of the Wannier configurations. For the $C_{4}$ symmetric case, when $n_{b}=2,n_{c}=0$, the corner-induced filling anomaly is $\eta^{(4)}=2$, when $n_{b}=n_{c}=1$, the corner-induced filling anomaly is $\eta^{(4)}=3,$ and when $n_{b}=0,n_{c}=2$, the corner-induced filling anomaly is $\eta^{(4)}=0$. All possible $C_{4}$ symmetric second-order HOTIs (atomic insulators and fragile TIs with zero bulk polarization) can be decomposed into these three basic cases. Thus, we can write the corner-induced filling anomaly as 

\begin{equation}
\label{eq:c4fa}
\eta^{(4)}=-n_{b}\ \rm{mod} \ 4.
\end{equation}
Through the same method, we can write the corner-induced filling anomalies of the $C_{2}$, $C_{3}$, and $C_{6}$ symmetric second-order HOTIs as
\begin{equation}
\label{eq:c236fa}
\begin{array}{l}
\eta^{(2)}=-n_{b} \ \rm{mod} \ 2,
\\
\eta^{(3)}=-n_{b} \ \rm{mod} \ 3,
\\
\eta^{(6)}= (2n_b + 3n_c)\ \rm{mod}\ 6 .
\end{array}
\end{equation}

Now we want to connect the corner-induced filling anomaly to the ES. According to Eq.\eqref{eq:c4fa}, to derive the corner-induced filling anomaly from the ES, we need to relate $n_{b}\ \mathrm{mod} \ 4$ with the ES. As discussed in Sec. \ref{subsec:ES}, if we put the rotation center at Wyckoff position $b$ and have cuts $A^{1}_{1/4}$ and $A^{2}_{1/2}$ in a $2L\times 2L$ ($L\rightarrow \infty$) square lattice, we can relate $\{n_{b}^{l}\}$ with the number of protected in-gap states in the ES. For simplicity, let us sort ${n_{b}^{0},n_{b}^{1},n_{b}^{2},n_{b}^{3}}$ and write them as ${n_{b}^{first},n_{b}^{second},n_{b}^{third},n_{b}^{fourth}}$ such that $n_{b}^{first}\geqslant n_{b}^{second}\geqslant n_{b}^{third}\geqslant n_{b}^{fourth}$. Then, we have $n_{b}=\sum_{l}n^{l}_{b}=n_{b}^{first}+n_{b}^{second}+n_{b}^{third}+n_{b}^{fourth}$. Using Eqs.\eqref{eq:cwops}, \eqref{eq:onehalf}, and \eqref{eq:onefourth}, we have 
\begin{equation}
\label{eq: numberWO1}
\begin{aligned}
    \mathrm{no}.\left[1/4,3/4\right]_{p,4,1}^{(4)}&=n_{b}^{first}-n_{b}^{fourth} \\
    \mathrm{no}.\frac{1}{4}_{p,4,1}^{(4)}&=n_{b}^{first}-n_{b}^{second} \\
    \mathrm{no}.\frac{1}{2}_{p,2,2}^{(4)}&=n_{b}^{first}-n_{b}^{second}+n_{b}^{third}-n_{b}^{fourth} \textbf{ or } \\
    &=n_{b}^{first}-n_{b}^{third}+n_{b}^{second}-n_{b}^{fourth},
\end{aligned}
\end{equation}
from which we can derive
\begin{equation}
\label{eq: numberWO2}
    n_{b}^{second}-n_{b}^{third}=\left|\mathrm{no}.\frac{1}{2}_{p,2,2}^{(4)}-\mathrm{no}.\left[1/4,3/4\right]_{p,4,1}^{(4)}\right|.
\end{equation}
This finally allows us to write down the number of orbitals located at Wyckoff position b as:
\begin{equation}
\label{eq:c4esnb}
\begin{aligned}
n_{b}&=n_{b}^{first}+n_{b}^{second}+n_{b}^{third}+n_{b}^{fourth}
\\
&=4n_{b}^{first}-2(n_{b}^{first}-n_{b}^{second})
\\
&-(n_{b}^{second}-n_{b}^{third})-(n_{b}^{first}-n_{b}^{fourth})
\\
&=4n_{b}^{first}-2\mathrm{no}.\frac{1}{4}_{p,4,1}^{(4)}-\mathrm{no}.\left[1/4,3/4\right]_{p,4,1}^{(4)}
\\
&-\left|\mathrm{no}.\frac{1}{2}_{p,2,2}^{(4)}-\mathrm{no}.\left[1/4,3/4\right]_{p,4,1}^{(4)}\right|.
\end{aligned}
\end{equation}
Note that in the last step of the above equation, we used Eqs.\eqref{eq: numberWO1} and \eqref{eq: numberWO2}. From Eqs.\eqref{eq:c4fa} and \eqref{eq:c4esnb}, we can write the corner-induced filling anomaly in terms of the number of the protected in-gap states as
\begin{equation}
\label{eq:c4esfa}
\begin{aligned}
\eta^{(4)}=&2\mathrm{no}.\frac{1}{4}_{p,4,1}^{(4)}+\mathrm{no}.\left[1/4,3/4\right]_{p,4,1}^{(4)}
\\
&+\left|\mathrm{no}.\frac{1}{2}_{p,2,2}^{(4)}-\mathrm{no}.\left[1/4,3/4\right]_{p,4,1}^{(4)}\right| \ \rm{mod} \ 4.
\end{aligned}
\end{equation}

Next, using the relationships between the ES and the configuration of Wannier orbitals in $C_{2}$, $C_{3}$, and $C_{6}$ symmetric second-order HOTIs (shown in Appendix \ref{app:CnDeriv}), we derive the corner-induced filling anomalies through the same method.
\begin{equation}
\label{eq:c236esfa}
\begin{array}{l}
\begin{aligned}
\eta^{(2)}=\mathrm{no}.\frac{1}{2}_{p,2,1}^{(2)} \rm{mod} \ 2,
\end{aligned}
\\
\\
\begin{aligned}
\eta^{(3)}=\mathrm{no}.\frac{1}{3}_{p,3,1}^{(3)}+\mathrm{no}.\left[1/3, 2/3\right]_{p,3,1}^{(3)} \ \rm{mod} \ 3,
\end{aligned}
\\
\\
\begin{aligned}
\eta^{(6)}=&-2\left(\mathrm{no}.\frac{1}{3}_{p,3,1}^{(6)}+\mathrm{no}.\left[1/3,2/3\right]_{p,3,1}^{(6)} \right)
\\
&+3\mathrm{no}.\frac{1}{2}_{p,2,1}^{(6)}\ \rm{mod} \ 6.
\end{aligned}
\end{array}
\end{equation}
The details of the derivations are shown in Appendix \ref{app:CnDeriv}. In conclusion, utilizing the relationships in Sec. \ref{subsec:ES}, we express the corner-induced filling anomaly, which captures the  quantized fractional corner charges in $C_n$ symmetric insulators, in terms of the number of \emph{protected} in-gap states in the ES for several different symmetric cuts. This is one of the main results of this paper.

\subsection{The role of interactions}
\label{subsec: int}
As discussed in Refs. \onlinecite{else2019fragile,li2019fractional}, if a set of Wannier orbitals at the same location forms a reducible representation that is the direct sum of all possible one-dimensional irreps of the $C_{n}$ group, no matter if there are interactions or not, we can always adiabatically move them from one maximal Wyckoff position to another through the (continuously deformable) general Wyckoff positions without breaking the $C_{n}$ rotation symmetry. 
Since this kind of deformation changes all $n_{\mathbf{q}}^{l}$ by $\pm 1$ at the same time, it cannot affect the difference between $n_{\mathbf{q}}^{l}$ with the same $\mathbf{q}$, and thus cannot affect the number of protected in-gap states in the ES according to the discussion in Sec. \ref{sec:ESWannier}. 

However, when interactions which preserve $C_{n}$ rotation symmetry and $U(1)$ charge conservation symmetry are present, we can adiabatically deform a set of Wannier orbitals in one unit cell with configuration $\{n_{\alpha}^{l}\}$ into another set of Wannier orbitals with configuration $\{(n_{\alpha}^{l})^{\prime}\}$, as long as the charge $Q\equiv \sum_{\alpha,l}M_{\alpha}n_{\alpha}^{l}$ and the angular momentum $L\equiv (\sum_{\alpha,l}lM_{\alpha}n_{\alpha}^{l}) \mod n$ are invariant. This kind of deformation \emph{can change} the number of protected in-gap states in the ES. 

To illustrate the idea, let us study a specific example having $C_{4}$ symmetry. 
we start from a (non-interacting) TCI with the configuration of Wannier orbitals $\{n_{b}^{1}=n_{b}^{3}=1, \mathrm{others}=0\}$, and continuously deform it into another (non-interacting) TCI with the configuration of Wannier orbitals $\{n_{b}^{2}=2,\mathrm{others}=0 \},$ which can be done in the presence of interactions during the deformation. According to Sec. \ref{subsec:ES}, the initial TCI has $\mathrm{no}.\frac{1}{4}_{p,4,1}^{(4)}=0$ and  $\mathrm{no}.\left[1/4,3/4\right]_{p,4,1}^{(4)}=1$ for the cut $A^{1}_{1/4}$, and has $\mathrm{no}.\frac{1}{2}_{p,2,2}^{(4)}=0$ for the cut $A^{2}_{1/2}$. After deforming $\{n_{b}^{1}=n_{b}^{3}=1, \mathrm{others}=0\}$ into $\{n_{b}^{2}=0, \mathrm{others}=0\}$ adiabatically, the final TCI has $\mathrm{no}.\frac{1}{4}_{p,4,1}^{(4)}=\mathrm{no}.\left[1/4,3/4\right]_{p,4,1}^{(4)}=2$ for the cut $A^{1}_{1/4}$, and $\mathrm{no}.\frac{1}{2}_{p,2,2}^{(4)}=2$ for the cut $A^{2}_{1/2}$, which are different from the initial values. We show more detailed calculations in Sec. \ref{subsec: intzcl}.

For this specific case, the change of the number of in-gap states in the single-particle ES for the cut $A^{1}_{1/4}$ and $A^{2}_{1/2}$ can be reflected in the change of the degeneracy of the many-body entanglement ground state in a certain sector (specified by the number of particles $N_A$ in region A of our entanglement bi-partition). This can be seen from the relationship between the many-body ES ($\{\zeta\}_{N_A}$) and the single-particle ES ($\{\xi_j\}$) \cite{alexandradinata2011trace}:
\begin{equation}
\label{eq:manybody}
\{\zeta\}_{N_{A}}=\left\{\prod_{i \in \text { occ }} \xi_{i} \prod_{j \in \text { unocc}}\left(1-\xi_{j}\right)\bigg| \sum_{i} n_{i}=N_{A}\right\},
\end{equation}
where $\{\zeta\}_{N_{A}}$ is the set of many-body entanglement eigenvalues of the $N_{A}$-particle sector of the many-body reduced density matrix $\rho_{A}$, and $n_{i}$ is the number of occupied states with eigenvalue $\xi_{i}$ in the single-particle ES. Eq. \eqref{eq:manybody} just represents filling the single-particle ES with $N_A$ particles to calculate the many-body ES. Filling the $N_A$ most entangled states in the single-particle ES will generate the many-body ES ground state. This illustrates that  degeneracy in the single particle ES can lead to  degeneracy in the many-body ES at particular fillings $N_A$. 

For some entanglement cuts the protected modes exist within a range of single-particle ES values, and are not fixed to lie at a single value. In these cases the change of the number of in-gap single-particle ES states in an interval for the cut $A^{m_{2}}_{1/m_{1}}$ corresponds to the change of the \emph{protected} quasi-degeneracy of the many-body entanglement ground state(s) in the $N_A=N(\xi>1-1/m_{1})+1$ sector, where $N(\xi>1-1/m_{1})$ is the number of states with eigenvalues $\xi>1-1/m_{1}$ in the single-particle ES for the cut $A^{m_{2}}_{1/m_{1}}$. In order to illustrate this idea, let us consider an ideal case where there are only protected in-gap states in the range of $\left[1/m_{1}, 1-1/m_{1}\right]$, i.e., any accidental states in this interval have been removed. According to Eq. \eqref{eq:manybody}, by filling all $N(\xi>1-1/m_{1})$ states with eigenvalues greater than $1-1/m_{1}$, and one of the protected in-gap states in the range of $\left[1/m_{1},1-1/m_{1}\right]$, we can have a 
subspace of size $\mathrm{no}.\left[1/m_{1}, 1-1/m_{1}\right]_{p,m_{1},m_{2}}$ that are protected quasi-degenerate eigenstates of $\rho_{A}$ in the $N_A=N(\xi>1-1/m_{1})+1$ sector. The state with the greatest eigenvalue is the absolute entanglement ground state of this sector, and the others are quasi-degenerate. Since the possible greatest and smallest eigenvalues are given by filling the protected in-gap states with eigenvalues $1-1/m_{1}$ and $1/m_{1}$, respectively, then the greatest eigenvalue is at most $(m_{1}-1)^2$ times larger than the smallest eigenvalue. Thus, the protected quasi-degenerate states are in the range of $\left[max/(m_{1}-1)^2, max\right]$, where $max$ is the greatest eigenvalue of the reduced density matrix $\rho_{A}$ in the $N_A=N(\xi>1-1/m_{1})+1$ sector. If there are accidental in-gap states in the range of $\left[1/m_{1}, 1-1/m_{1}\right]$, they only contribute to the \emph{accidental} quasi-degeneracy, which can be removed without breaking the symmetry or closing the bulk gap. Note that when $m_1=2$, the quasi-degeneracy becomes the exact degeneracy at $1/2$ discussed in Refs \cite{hughes2011inversion,fang2013entanglement}.

Even though the number of protected in-gap states changes after the deformation in the presence of interactions, the corner-induced filling anomaly given by Eq.\eqref{eq:c4esfa} is \emph{invariant} under this deformation. To see this we note that the allowed symmetric adiabatic deformation in the presence of interactions can only change $n^{l}_{\mathbf{q}}$'s but keeps $n_{\mathbf{q}}$ invariant \cite{li2019fractional}, and the corner-induced filling anomaly only depends on $n_{b}$ and $n_{c}$ as shown in Eqs.\eqref{eq:c4fa} and  \eqref{eq:c236fa}. In conclusion, the number of protected in-gap states in the ES is not robust when interactions are present, and its non-robustness leads to the non-robustness of the quasi-degeneracy of the many-body ES ground state subspace during continuous deformations when symmetric interactions are present. However, the corner-induced filling anomaly and the formulas for it in Eqs.\eqref{eq:c4esfa}, \eqref{eq:c236esfa}, \eqref{eq: fillingac4} and \eqref{eq:c236esfazll} are robust as long as the bulk gap and symmetry are preserved. 

\section{Zero correlation length limit}
\label{sec: zcl}
\begin{figure}[htp]
\centering
\includegraphics[width=1\columnwidth]{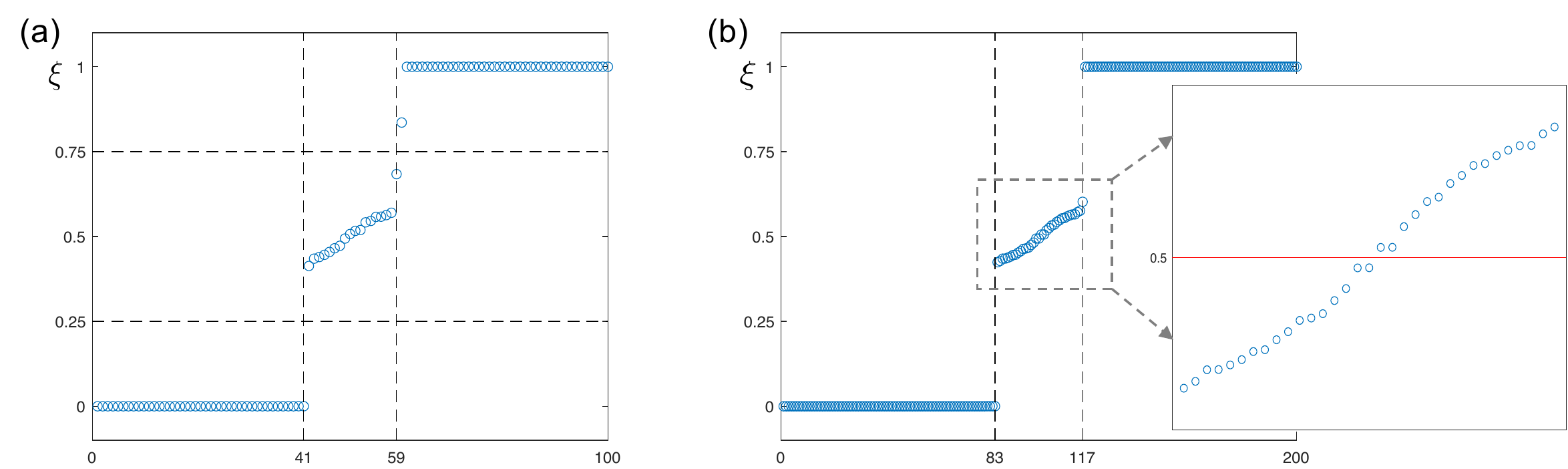}
\caption{Entanglement spectrum of $h^{(4)}_{2b}$ in Eq. \eqref{eq:h2b} with $\lambda_1=\lambda_2=0.15$ in a $10\times 10$ square lattice with open boundary conditions along both directions for cut (a)$A^{1}_{1/4}$ which exhibits  18 in-gap states in the range of $\left[1/4,3/4\right]$, and no in-gap state at $1/4$; and for cut (b)$A^{2}_{1/2}$ for which there are no in-gap states at $1/2.$ The x-axis represents the sequence number of eigenstates of the reduced correlation operator, and the y-axis represents the corresponding eigenvalues.}
\label{fig:h2bf}
\end{figure}

Up to now, we have established the generic relationships between the configuration of Wannier orbitals and the number of \emph{protected} in-gap states in the ES. However, with these relationships, we cannot simply obtain the correct filling anomaly of a generic Hamiltonian describing a TCI by directly calculating its ES for symmetric cuts because of the possible presence of accidental/unprotected modes. Here is an illustrative example: in Fig. \ref{fig:h2bf}, we numerically calculate the ES for cuts $A^{1}_{1/4}$ and $A^{2}_{1/2}$ in a $10\times 10$ open square lattice for the $h^{(4)}_{2b}$ model shown in Eq. \eqref{eq:h2b} ($n_{b}^{0}=n_{b}^{2}=0$, $n_{b}^{1}=n_{b}^{3}=1$, and $\mathrm{others}=0$).
For the parameters $\lambda_1=\lambda_2=0.15,$ and the $A^{1}_{1/4}$ cut,  we find 18 in-gap states with eigenvalues in the interval $\left[1/4,3/4\right]$, and no in-gap states with eigenvalue $\frac{1}{4}$ (as shown in Fig. \ref{fig:h2bf}(a)). Also, there are no in-gap states with eigenvalue $1/2$ in the $A^{2}_{1/2}$ cut (as shown in Fig. \ref{fig:h2bf}(b)). Substituting these results into Eq. \eqref{eq:c4esfa}, we find $\eta=0,$ not $2$ as we expected from Eq. \eqref{eq:c4fa}. This discrepancy is because \emph{not all} in-gap states of this model are \emph{protected}. In general, there can be accidental in-gap states that interfere with the identification of the protected in-gap states. Indeed, in principle we need to sample all possible continuous, symmetry-preserving deformations of a Hamiltonian to identify which in-gap states in its ES are protected. This will  not generically be numerically efficient, although it may be possible in some cases. 

To counter this issue we can instead try to adiabatically tune the model to the zero correlation length limit (ZCL), where the correlation length is zero. While this limit \emph{does not} remove all of the accidental in-gap states, it will allow us to refine our formulas so that we need not separate protected from unprotected in-gap ES states. Indeed the physical origin of all the in-gap states can be understood clearly and intuitively using the Wannier orbital picture thanks to the simplicity of the wavefunctions of Wannier orbitals in the ZCL. For this reason, in the ZCL we can express the filling anomaly in terms of the total number of in-gap states (both protected or accidental ones), which can be checked by straightforward numerical calculations. Thus, in this section we will focus our discussion on the ZCL which will help us gain more intuition about the relationships among the configuration of Wannier orbitals, the ES, and the filling anomaly.

\subsection{Physical origin of the in-gap states in the ES}
\label{subsec:zerol}
\begin{figure}[htp]
\centering
\includegraphics[width=1\columnwidth]{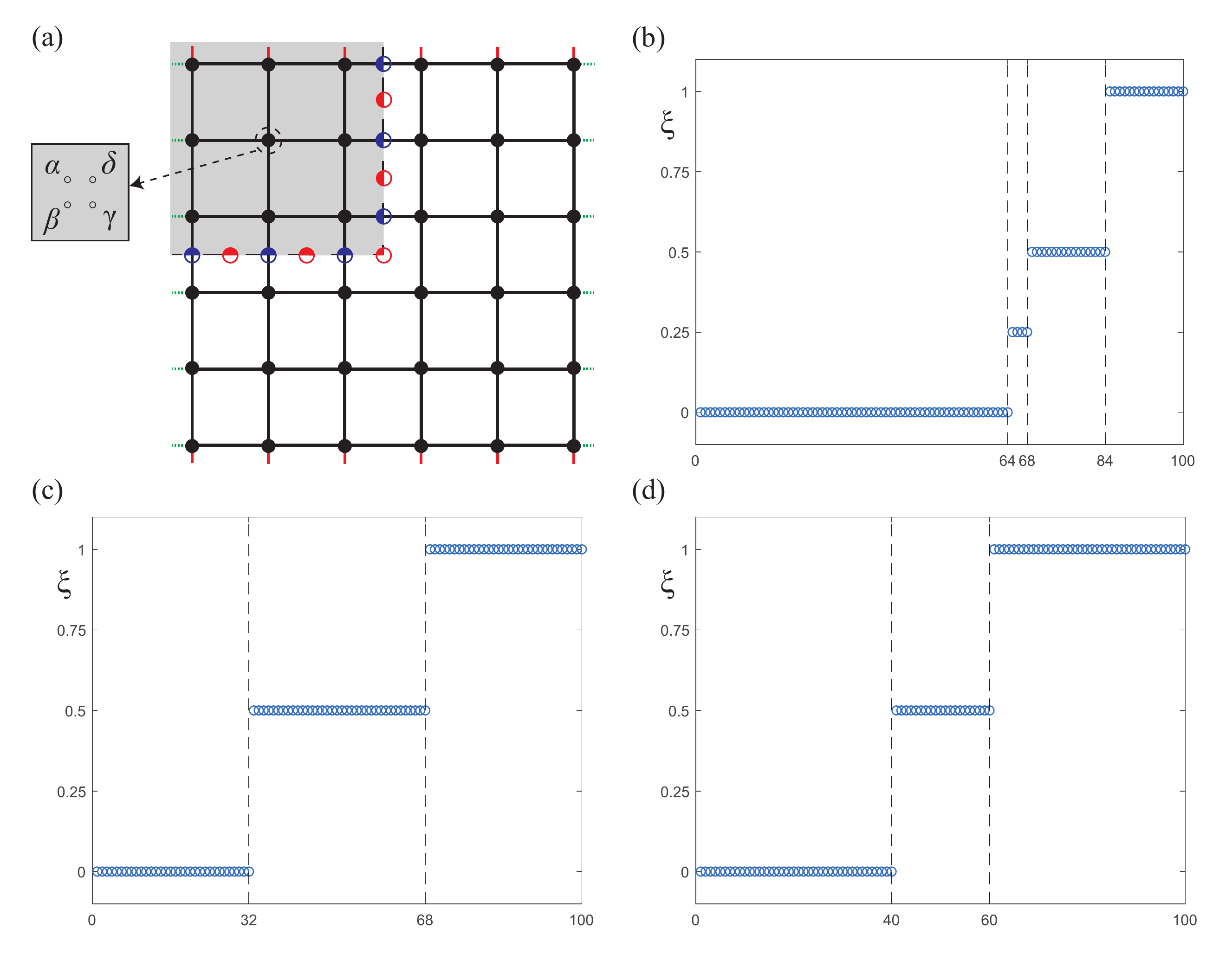}
\caption{(a) Schematic illustration of the corner cut in the periodic lattice. The red solid (green dotted) lines on edges indicate the hopping between top and bottom (right and left) edges. Wannier orbitals localized at $b$ and $c$ are represented by red dots and blue dots respectively. The solid parts of the dots indicate the parts of Wannier orbitals which are included in the subset $A$. The ES is shown for the cut $A^{1}_{1/4}$ shown in panel (a) in a $10\times 10$ square lattice with periodic boundary conditions for (b) $h^{(4)}_{1b}$, (c) $h^{(4)}_{2b}$, and (d)$h^{(4)}_{2c}$ models in the ZCL. The x-axis represents the sequence number of eigenstates of the reduced correlation operator, and the y-axis represents the corresponding eigenvalues.}
\label{fig:C4fourPlots}
\end{figure}

In order to understand the physical origin of the in-gap states, we need to first discuss the details of the wavefunctions of Wannier orbitals in the ZCL. Again, taking $C_{4}$ symmetric atomic insulators as an example, let us start from Wannier orbitals localized at Wyckoff position $b$. Without loss of generality, we can consider four internal degrees of freedom in one unit cell that are (cyclically) related by four-fold rotation symmetry (labelled by $\alpha$, $\beta$, $\gamma$ and $\delta$ in Fig. \ref{fig:C4fourPlots}(a)). From these degrees of freedom we can  generate representations of Wannier orbitals localized at Wyckoff position $b$ that span each possible angular momentum (i.e., each $C_4$ irrep). A simple construction generates:
\begin{equation}
\label{eq: wannierfuncb_detail}
\begin{array}{l}
\begin{aligned}
\ket{W_{b,0}}=\frac{1}{2}(&\ket{x,y,\delta}+\ket{x+1,y,\alpha}
\\
&+\ket{x+1,y+1,\beta}+\ket{x,y+1,\gamma}),
\end{aligned}
\\
\begin{aligned}
\ket{W_{b,1}}=\frac{1}{2}(&-i\ket{x,y,\delta}-\ket{x+1,y,\alpha}\\
&+i\ket{x+1,y+1,\beta}+\ket{x,y+1,\gamma}),
\end{aligned}
\\
\begin{aligned}
\ket{W_{b,2}}=\frac{1}{2}(&-\ket{x,y,\delta}+\ket{x+1,y,\alpha}
\\
&-\ket{x+1,y+1,\beta}+\ket{x,y+1,\gamma}),
\end{aligned}
\\
\begin{aligned}
\ket{W_{b,3}}=\frac{1}{2}(&i\ket{x,y,\delta}-\ket{x+1,y,\alpha}
\\
&-i\ket{x+1,y+1,\beta}+\ket{x,y+1,\gamma}),
\end{aligned}
\end{array}
\end{equation}
where the subscript $b,l$ denotes Wannier orbitals with angular momentum $l$ at Wyckoff position $b$. Thus, these four degrees of freedom can be used to generate one Wannier orbital at $b$ with each angular momentum. If we want to have two Wannier orbitals at $b$ with the same angular momentum, we can use another set of four internal degrees of freedom $\alpha^{\prime}$, $\beta^{\prime}$, $\gamma^{\prime}$ and $\delta^{\prime}$ to generate another complete set of Wannier orbitals with angular momentum $l=0,1,2,3$ ($\ket{W_{b,l}^{\prime}}$). In this way, we can construct collections of wavefunctions representing any configuration of Wannier orbitals at $b$ (i.e., any $\{n_{b}^{l}\}$) that we want. 

Next, let us consider the ES for the $A^{1}_{1/4}$ cut. If we have one Wannier orbital at $b$ filled, i.e., $n_{b}=\sum_{l}n^{l}_{b}=1$, the reduced correlation operator is
\begin{equation}
\label{eq: corrzcl}
C_{A}=P_{A}\sum_{\mathbf{R}}\ket{W_{b,l}(\mathbf{R})}\bra{W_{b,l}(\mathbf{R})}P_{A},
\end{equation}
where $P_{A}$ is the projector on region $A$ (shaded region in Fig. \ref{fig:C4fourPlots}(a)), and $\mathbf{R}$ labels the unit cell. As shown in Fig. \ref{fig:C4fourPlots}(a), the Wannier orbital at the corner of the cut  has  only one quarter of its density in region $A$. More specifically, it contributes a term $\frac{1}{4}\ket{\frac{L}{2},\frac{L}{2}+1,\gamma}\bra{\frac{L}{2},\frac{L}{2}+1,\gamma}$ to the correlation operator, where $L$ is the number of unit cells along one direction of the square lattice. Hence, this term leads to an in-gap state with eigenvalue $1/4$ in the ES. By similar arguments, a Wannier orbital at $b$ that is localized on the edges of the cut contributes an in-gap state with eigenvalue $1/2$ in the ES. In Fig. \ref{fig:C4fourPlots}(b), we numerically calculate the ES for the cut $A^{1}_{1/4}$ in a $10\times 10$ periodic square lattice for the $h^{(4)}_{1b}$ model ($n^{l}_{b}=\delta_{l,2}$ for $l=0,1,2,3$)\cite{benalcazar2018quantization}: 
\begin{equation}
\label{eq:h1b}
\begin{aligned}
h^{(4)}_{1b}(\mathbf{k})=&e^{ik_{y}}\left(\hat{c}^{\dag}_{1,\mathbf{k}}\hat{c}_{2,\mathbf{k}}+\hat{c}^{\dag}_{4,\mathbf{k}}\hat{c}_{3,\mathbf{k}}\right)
\\
&+e^{ik_{x}}\left(\hat{c}^{\dag}_{3,\mathbf{k}}\hat{c}_{2,\mathbf{k}}+\hat{c}^{\dag}_{4,\mathbf{k}}\hat{c}_{1,\mathbf{k}}\right)+h.c.,
\end{aligned}
\end{equation}
where $c_{a,\mathbf{k}}$ is the Fourier transformation of $c_{a,\mathbf{R}}$, which is the annihilation operator of a spinless electron on sublattice $a$  in the unit cell located at $\mathbf{R}$ on a square lattice. 
The size of the cut region is $5\times 5$ and we find four states with eigenvalue $1/4$ from the four corners, and 16 states with eigenvalue $1/2$ from the four edges (four states on each edge of the cut), which is consistent with our predictions. 

Now let us consider $n_{b}=2.$ In this case, at the corners of the cut, we could either have one in-gap state with eigenvalue $1/2,$ or two in-gap states with eigenvalue $1/4$ depending on whether these two Wannier orbitals are generated by an overlapping set of internal degrees of freedom (four internal degrees of freedom related by $C_{4}$ rotation) or not. If they are generated by the same set of internal degrees of freedom (e.g., the functions $\ket{W_{b,1}}$ and $\ket{W_{b,2}}$ listed above), then we call them a doublet, and they contribute a term $\frac{1}{2}\ket{\frac{L}{2},\frac{L}{2}+1,\gamma}\bra{\frac{L}{2},\frac{L}{2}+1,\gamma}$ to the correlation operator, which leads to one in-gap state with eigenvalue $1/2$.  If they are generated by orthogonal degrees of freedom (e.g., the functions $\ket{W_{b,1}}$ and $\ket{W_{b,1}^{\prime}}$ listed above), we call them two singlets, and they contribute the terms $\frac{1}{4}\ket{\frac{L}{2},\frac{L}{2}+1,\gamma}\bra{\frac{L}{2},\frac{L}{2}+1,\gamma}+\frac{1}{4}\ket{\frac{L}{2},\frac{L}{2}+1,\gamma^{\prime}}\bra{\frac{L}{2},\frac{L}{2}+1,\gamma^{\prime}}$ to the correlation operator. This leads to two in-gap states with eigenvalue $1/4$.

We can similarly define triplets and quadruplets if we have more Wannier orbitals. Generally, an atomic insulator with non-zero $n_{b}$ can be represented as a combination of singlets, doublets, triplets, and quadruplets, which contribute corner states with ES eigenvalues $1/4$, $1/2$, $3/4$ and $1$, respectively. By similar arguments, singlets contribute edge states with eigenvalue $1/2$. Doublets contribute a pair of edge states either both with eigenvalue $1/2$ (for example, $n_{b}^{0}=n_{b}^{2}=1$ and $n_{b}^{1}=n_{b}^{3}=0$), or a pair with the eigenvalues $(2\pm\sqrt{2})/4$ (for example, $n_{b}^{0}=n_{b}^{1}=1$ and $n_{b}^{2}=n_{b}^{3}=0$). This is because for $n_{b}^{0}=n_{b}^{2}=1$ and $n_{b}^{1}=n_{b}^{3}=0$, we can get a term $\frac{1}{2}\left(\ket{x,y,\delta}\bra{x,y,\delta}+\ket{x,y+1,\gamma}\bra{x,y+1,\gamma}\right)$ in the reduced correlation operator, which leads to a pair of in-gap states both with eigenvalue $1/2$; and for $n_{b}^{0}=n_{b}^{1}=1$ and $n_{b}^{2}=n_{b}^{3}=0$, we can get a term $\frac{1}{4}[2\ket{x,y,\delta}\bra{x,y,\delta}+2\ket{x,y+1,\gamma}\bra{x,y+1,\gamma}+(1-i)\ket{x,y,\delta}\bra{x,y+1,\gamma}+(1+i)\ket{x,y+1,\gamma}\bra{x,y,\delta}]$, which leads to a pair of in-gap states with eigenvalue $(2\pm\sqrt{2})/4$.  Triplets contribute a pair of edge states with eigenvalues $1/2$ and $1$, respectively, and quadruplets contribute a pair of edge states each with eigenvalue $1$. In Fig. \ref{fig:C4fourPlots}(c), we numerically calculate the ES for the cut $A^{1}_{1/4}$ in a $10\times 10$ periodic square lattice for the $h^{(4)}_{2b}$ model ($n_{b}^{0}=n_{b}^{2}=0$ and $n_{b}^{1}=n_{b}^{3}=1$) in Eq. \eqref{eq:h2b} with $\lambda_1=\lambda_2=0$. There are four states with eigenvalue $1/2$, one from each of the four corners, and 32 states with eigenvalue $1/2$ with eight coming from each of the four edges, which is consistent with our predictions.

Next, let us consider Wannier orbitals localized at $c$. Note that in our construction the contributions to in-gap states in the ES from Wannier orbitals localized at $b$ and $c$ are independent. Since each Wannier orbital localized at $c$ is only $C_{2}$ symmetric, its angular momentum should be either $0$ or $1$. Following the same method in the previous paragraphs, we conclude that singlets of Wannier orbitals localized at $c$ on the edges of the cut contribute in-gap states with eigenvalue $1/2$ in the ES, while doublets do not contribute in-gap states. More importantly, Wannier orbitals localized at $c$ \emph{do not} contribute corner states in the cut $A_{1/4}^{1}$, because the Wyckoff position $c$ can never be the corner of the cut $A_{1/4}^{1}$.  As shown in Fig. \ref{fig:C4fourPlots}(d), we numerically calculate the ES for the cut $A^{1}_{1/4}$ in a $10\times 10$ periodic square lattice for the $h^{(4)}_{2c}$ model ($n^{1}_{c}=1$ and $n^{0}_{c}=0$)\cite{benalcazar2018quantization}:
\begin{equation}
h^{(4)}_{2c}(\mathbf{k})=e^{ik_{x}}\hat{c}^{\dag}_{3,\mathbf{k}}\hat{c}_{1,\mathbf{k}}
+e^{ik_{y}}\hat{c}^{\dag}_{4,\mathbf{k}}\hat{c}_{2,\mathbf{k}}+h.c.,
\end{equation}
where $c_{a,\mathbf{k}}$ is the Fourier transformation of $c_{a,\mathbf{R}}$, which is the annihilation operator of a spinless electron on sublattice $a$  in the unit cell located at $\mathbf{R}$ on a square lattice. There are 20 states with eigenvalues $1/2$ with five coming from each of the four edges, which is consistent with our predictions. 

Finally, since Wannier orbitals localized at Wyckoff position $a$ (the unit cell center) can never be divided by the cut as long as the cut does not break the unit cell, there are no in-gap states from Wannier orbitals localized at this position. Note that these simple relationships between the (total) number of in-gap states in the ES, and configuration of Wannier orbitals are only applicable to atomic insulators, because fragile TIs have no zero correlation length limit. In conclusion, given an arbitrary Hamiltonian describing an atomic insulator in the zero correlation length limit, the total number of in-gap states in the ES for different symmetric cuts, including both protected and accidental ones, can be determined from the configuration of Wannier orbitals. For $C_{2}$, $C_{3}$, and $C_{6}$ symmetric atomic insulators, we derive similar conclusions in the zero correlation length limit through the same method, for which details are shown in Appendix \ref{app:CnDeriv}.

\subsection{Filling Anomaly and ES in the Zero Correlation Length Limit}
\label{subsec: fazcl}
\begin{figure*}
\centering
\includegraphics[width=2\columnwidth]{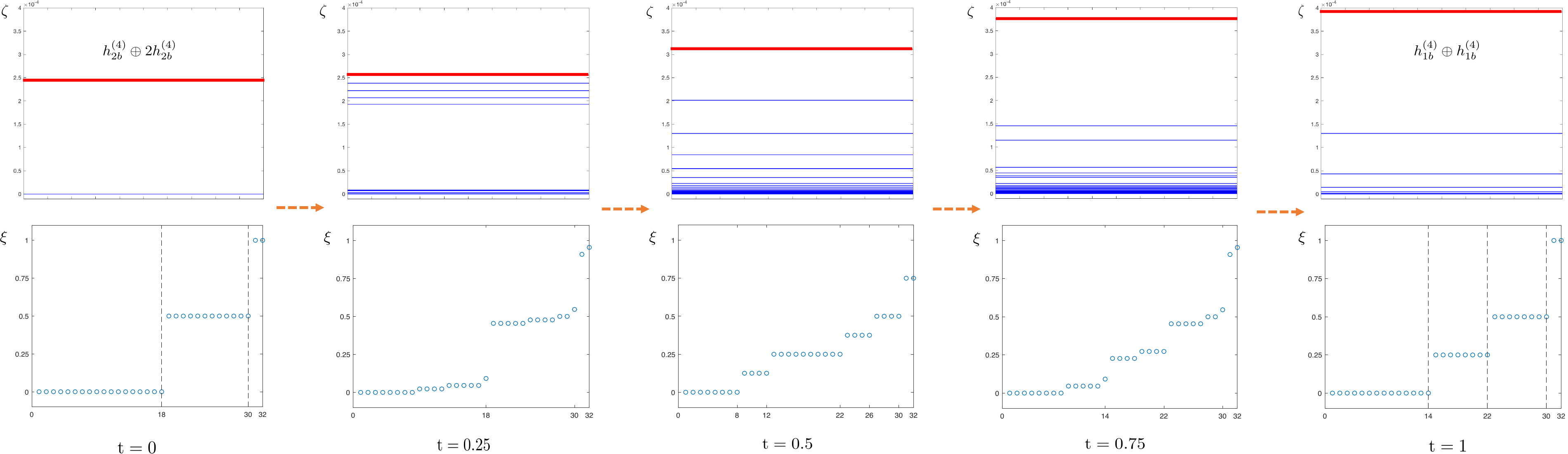}
\caption{\label{fig:int}The upper five pictures are the spectra (all the eigenvalues for all values of $N_A$) of the reduced density matrix $\rho_{A^{1}_{1/4}}$,$\{\zeta\}$, in a $4\times 4$ periodic square lattice for models described by Eq.\eqref{eq:int}, and the lower five pictures correspond to the spectra of the single-particle correlation functions. The x-axis represents the sequence number of eigenstates of the reduced correlation operator, and the y-axis represents the corresponding eigenvalues. At $t=0$ and $1$, we have higher-order TCIs $h^{(4)}_{2b}\oplus 2h^{(4)}_{2b}$ and $h^{(4)}_{1b}\oplus h^{(4)}_{1b}$, respectively, both at $1/4$ filling. The degeneracy of the entanglement ground state (marked with red lines) changes from $4096$ to $256$ when we turn on the interactions and tune $t$ from $0$ to $1$.}
\end{figure*}
Using our direct relationship between the Wannier configurations and the in-gap ES modes, we can derive simplified formulas for the filling anomaly that apply (at least) in the zero-correlation length limit. For example, in $C_{4}$ symmetric cases, we can represent the corner-induced filling anomaly in terms of the number of in-gap states (both protected and accidental) in the ES for the cut $A^{1}_{1/4}$ in a square lattice with periodic boundary conditions. From the discussion in Sec. \ref{subsec:ES}, we know that the corner states in the ES tell us the information about $n_{b}$:
\begin{equation}
n_{b}\ \rm{mod} \ 4=\frac{1}{4}\left(\mathrm{no}.\frac{1}{4}^{(4)}+2\mathrm{no}.\frac{1}{2}_{corner}^{(4)}+3\mathrm{no}.\frac{3}{4}^{(4)}\right) \ \rm{mod} \ 4,
\end{equation}
where $\mathrm{no}.\frac{j}{4}^{(4)}, j=1,2,3$ is the total number of in-gap corner states with eigenvalue $j/4$ in the ES for the $A_{1/4}^{1}$ cut in the $C_{4}$ symmetric case, and the subscript ``corner" of $\mathrm{no}.\frac{1}{2}^{(4)}$ is to emphasize that we are referring to the corner states with eigenvalue $1/2$, since the states with eigenvalue $1/2$ come from both the edges and the corners of the cut. Note that the overall factor of $1/4$ is because in the periodic square lattice, the $A_{1/4}^{1}$ cut has four corners. 

To derive the corner-induced filling anomaly, we need to calculate the parity of $\frac{1}{4}\left(\mathrm{no}.\frac{1}{2}\right)_{corner}^{(4)},$ i.e., we do not need its full value. It is not obvious that this quantity is simple to isolate since it requires us to separately consider the edge and corner contributions that generate the 1/2 eigenvalues, however, we will now show that the parity of $\frac{1}{4}\left(\mathrm{no}.\frac{1}{2}\right)_{corner}^{(4)}$ can be recast in terms of \emph{all} the $1/2$ eigenvalues, no matter their origin. First we note that the quantity $\frac{1}{4}\mathrm{no}.\frac{1}{4}^{(4)}$ tells us the number of singlets of Wannier orbitals at $b,$ and  $\frac{1}{4}\mathrm{no}.\frac{3}{4}^{(4)}$ tells us the number of triplets of Wannier orbitals at $b$. Also, we denote the number of in-gap states from edges \emph{and} corners with eigenvalue $1/2$ as $\mathrm{no}.\frac{1}{2}^{(4)}$. Since the cut $A_{1/4}^{1}$ has four edges and four corners in the periodic square lattice, $\mathrm{no}.\frac{1}{2}^{(4)}$ must be a multiple of 4. According to Sec. \ref{sec:ESWannier}, $\mathrm{no}.\frac{1}{2}^{(4)}$ has contributions from singlets, doublets, and triplets of Wannier orbitals at $b$, and singlets of Wannier orbitals at $c$. Thus, 
\begin{equation}
\label{eq: onehalf0}
\begin{aligned}
\frac{1}{4}&\mathrm{no}.\frac{1}{2}^{(4)}=\frac{1}{4}(L_{cut}-1)(\mathrm{no}.\frac{1}{4}^{(4)}+\mathrm{no}.\frac{3}{4}^{(4)})
\\
&+L_{cut}n_{c,s}+2(L_{cut}-1)n_{b,d}^{\prime}
+\frac{1}{4}\left(\mathrm{no}.\frac{1}{2}^{(4)}\right)_{corner},
\end{aligned}
\end{equation}
where $L_{cut}$ is the length of edges of the cut $A_{1/4}^{1}$, $n_{c,s}$ is the number of singlet orbitals at Wyckoff position $c$, and $n_{b,d}^{\prime}$ is the number of doublet orbitals at Wyckoff position $b$ that contribute a pair of in-gap edge states with eigenvalue $1/2$. From this equation, and the requirement of vanishing bulk polarization, it can be shown that the parity of $\frac{1}{4}\mathrm{no}.\frac{1}{2}^{(4)}_{corner}$ is equal to the parity of $\frac{1}{4}(\mathrm{no}.\frac{1}{2}^{(4)}-\mathrm{no}.\frac{1}{4}^{(4)}-\mathrm{no}.\frac{3}{4}^{(4)})$ (details are shown in Appendix \ref{app:parity}). Thus, 
\begin{equation}
\label{eq: bes}
\begin{aligned}
n_{b}\ \rm{mod} \ 4&=\frac{1}{4}\left(-\mathrm{no}.\frac{1}{4}^{(4)}+2\mathrm{no}.\frac{1}{2}^{(4)}+\mathrm{no}.\frac{3}{4}^{(4)}\right) \ \rm{mod} \ 4
\\
&=-\frac{1}{4}\left(\mathrm{no}.\frac{1}{4}^{(4)}+2\mathrm{no}.\frac{1}{2}^{(4)}+3\mathrm{no}.\frac{3}{4}^{(4)}\right) \ \rm{mod} \ 4.
\end{aligned}
\end{equation}
Note that in the second step of the above equation, we used the fact that $\mathrm{no}.\frac{j}{4},j=1,2,3$ are always multiples of 4 because of the four-fold rotation symmetry. Utilizing  Eq. \eqref{eq: bes}, we finally arrive at the corner-induced filling anomaly:
\begin{equation}
\label{eq: fillingac4}
\eta^{(4)}=\frac{1}{4}\left(\mathrm{no}.\frac{1}{4}^{(4)}+2\mathrm{no}.\frac{1}{2}^{(4)}+3\mathrm{no}.\frac{3}{4}^{(4)}\right) \ \rm{mod} \ 4.
\end{equation} An important feature of this final result is that it does not depend on the separate evaluation of edge and corner modes at $1/2,$ and instead requires only the total number of modes at the relevant quantized eigenvalues.

By the same method, we can write the corner-induced filling anomaly for $C_n$ symmetric insulators in terms of the number of in-gap states with different quantized eigenvalues in the ES of the $A^{1}_{1/n}$ cut with $n=2,3,6$:
\begin{equation}
\label{eq:c236esfazll}
\begin{array}{l}
\begin{aligned}
\eta^{(2)} = \frac{1}{4}\left(\mathrm{no}.\frac{1}{2}^{(2)}\right)\ \rm{mod}\ 2,
\end{aligned}
\\
\\
\begin{aligned}
\eta^{(3)}= \frac{1}{3}\left(\mathrm{no}.\frac{1}{3}^{(3)}-\mathrm{no}.\frac{2}{3}^{(3)}\right) \ \rm{mod} \ 3,
\end{aligned}
\\
\\
\begin{aligned}
\eta^{(6)}=\frac{1}{6}\left(2\mathrm{no}.\frac{1}{3}^{(6)}-2\mathrm{no}.\frac{2}{3}^{(6)}+3\mathrm{no}.\frac{1}{2}^{(6)}\right) \ \rm{mod} \ 6.
\end{aligned}
\end{array}
\end{equation}
The details of the derivation of Eq. \eqref{eq:c236esfazll} are shown in Appendix \ref{app:CnDeriv}. In the zero correlation length limit, the in-gap states can take only quantized values in the ES for the cut $A^{1}_{1/n}$ in $C_{n}$ symmetric atomic insulators. Thus, we can simply read off these numbers from the ES, and then easily find the corner-induced filling anomaly using Eqs. \eqref{eq: fillingac4} or \eqref{eq:c236esfazll}.

\subsection{Interactions and the Zero Correlation Length Limit}
\label{subsec: intzcl}
In this subsection, we numerically study the evolution of the ES for the cut $A^{1}_{1/4}$ in the ZCL during the adiabatic deformation discussed in Sec \ref{subsec: int} to gain a better intuition. According to Sec. \ref{subsec:zerol}, we should have $\mathrm{no}.\frac{1}{2}^{(4)}=12$ and $\mathrm{no}.\frac{1}{4}^{(4)}=0$ before the deformation, and $\mathrm{no}.\frac{1}{2}^{(4)}=\mathrm{no}.\frac{1}{4}^{(4)}=8$ after the deformation. In Fig. \ref{fig:int}, we numerically calculate the full many-body entanglement spectrum $\{\zeta\}$ of the reduced density matrix $\rho_{A^{1}_{1/4}}$ in a $4\times 4$ periodic square lattice for the interacting Hamiltonian 
\begin{equation}
\label{eq:int}
\begin{aligned}
H&(\mathbf{k},t)=(1-t)h^{(4)}_{2b}(\mathbf{k})\oplus 2h^{(4)}_{2b}(\mathbf{k})+t h^{(4)}_{1b}(\mathbf{k})\oplus h^{(4)}_{1b}(\mathbf{k})
\\
&+\sin\left[(1-t)\pi\right]\left(\chi_{b,1}^{\dag}(\mathbf{k})\chi_{b,3}^{\dag}(\mathbf{k})\chi_{b,2}(\mathbf{k})\chi_{b,2}^{\prime}(\mathbf{k})+h.c.\right),
\end{aligned}
\end{equation}
where $h^{(4)}_{2b}(\mathbf{k})$ and $h^{(4)}_{1b}(\mathbf{k})$ are shown in Eqs. \eqref{eq:h2b} and \eqref{eq:h1b} respectively. The operators $\chi_{b,1}^{\dag}(\mathbf{k})$ and $\chi_{b,3}^{\dag}(\mathbf{k})$ are the creation operators of corresponding eigenstates of the lowest two bands of $h^{(4)}_{2b}\oplus 2h^{(4)}_{2b}$, and $\chi_{b,2}(\mathbf{k})$ and $\chi_{b,2}^{\prime}(\mathbf{k})$ are the creation operators of corresponding eigenstates of the lowest two bands of $h^{(4)}_{1b}(\mathbf{k})\oplus h^{(4)}_{1b}(\mathbf{k})$. It is clear from Eq. \eqref{eq:int} that $U(1)$ charge conservation is preserved. Additionally, since $C_{4}\chi_{b,l}^{\dag}(\mathbf{k})C_{4}^{-1}=e^{il\pi/2}\chi_{b,l}^{\dag}(R_{4}\mathbf{k})$, the quartic interaction term (the last term) in Eq. \eqref{eq:int} satisfies $C_{4}\chi_{b,1}^{\dag}(\mathbf{k})\chi_{b,3}^{\dag}(\mathbf{k})\chi_{b,2}(\mathbf{k})\chi_{b,2}^{\prime}(\mathbf{k})C_{4}^{-1}=\chi_{b,1}^{\dag}(R_{4}\mathbf{k})\chi_{b,3}^{\dag}(R_{4}\mathbf{k})\chi_{b,2}(R_{4}\mathbf{k})\chi_{b,2}^{\prime}(R_{4}\mathbf{k})$, which means the Hamiltonian in Eq. \eqref{eq:int} is $C_{4}$-symmetric at every $t$ during the deformation.

For the deformation we start at $t=0$ where there are eight bands for $H(t=0)=h^{(4)}_{2b}\oplus 2h^{(4)}_{2b}.$ We fill the lowest two of these bands which generates the Wannier configuration $\{n_{b}^{1}=n_{b}^{3}=1, \mathrm{others}=0\}$. As $t$ goes from $0$ to $1$, we continuously deform $H(t=0)=h^{(4)}_{2b}(\mathbf{k})\oplus 2h^{(4)}_{2b}(\mathbf{k})$ to $H(t=1)=h^{(4)}_{1b}(\mathbf{k})\oplus h^{(4)}_{1b}(\mathbf{k}).$ For the final system at $t=1$ the lowest two bands are again filled as the configuration of Wannier orbitals is $\{n_{b}^{2}=2, \mathrm{others}=0\}$. Thus, as $t$ goes from $0$ to $1$, we continuously deform $\{n_{b}^{1}=n_{b}^{3}=1, \mathrm{others}=0\}$ into $\{n_{b}^{2}=2, \mathrm{others}=0\}$ without closing the gap between the many-body ground state and the first excited state, and without breaking the symmetry.  During this process the degeneracy of the many-body entanglement ground state (which is marked with red lines in Fig. \ref{fig:int}) changes from $4096$ to $256$. This is expected from our calculations in the ZCL as it reflects the change of the number of in-gap states in the single-particle ES. 
In this specific example, by directly substituting the configurations of Wannier orbitals before and after the deformation into Eq. \eqref{eq: fillingac4}, we can see that the corner-induced filling anomaly remains invariant even though the number of in-gap states of the ES changes.

\section{Conclusion and discussion}
\label{sec: con}
In our study of two-dimensional $C_n$ symmetric second-order HOTIs, we have shown that the ordinary single-particle ES (as opposed to the nested ES) can be used to characterize the higher-order features including the corner-induced filling anomaly. More specifically, we expressed the corner-induced filling anomaly in terms of the number of (protected) in-gap states of the ES for various $C_n$ symmetric cuts. While there may be some features that are better diagnosed using the nested ES, the higher-order features characterized by the corner-induced filling anomaly, e.g., the observable fractional corner charge, can be determined from the ES of atomic and fragile insulators. Furthermore, we went on to characterize the properties of insulators with interactions, and also carefully studied the zero-correlation length limit. We found that even though the number of protected single-particle entanglement modes, and the corresponding many-body entanglement ground state degeneracy, are not invariant under symmetric adiabatic deformations in the presence of interactions, the corner-induced filling anomaly is invariant, and can be determined from the ES.

It is natural to ask if there are HOTI properties that are beyond the (first-order) ES and may require nested versions of the ES. For this to play an important role it is likely that one would need to move beyond the simple cyclic $C_n$ symmetry groups and include reflection symmetries. We leave a more detailed study to future work.


\section*{Acknowledgements}
PZ thanks Tianhe Li for useful discussions. PZ was supported by the National Science Foundation
Emerging Frontiers in Research and Innovation NewLAW program (grant
EFMA-1641084). KL was supported by the National Science Foundation under Grant PHY-1659598. TLH thanks the US National Science Foundation MRSEC program under NSF Award Number DMR-1720633 (iSuperSEED) for
support. TLH also thanks the National Science Foundation under Grant No. NSF PHY-1748958(KITP) for partial support at the end stage of this work during the Topological Quantum Matter program.

\bibliography{apssamp}

\appendix
\section{\label{app:review}Review on key conclusions in Ref \onlinecite{fang2013entanglement}}
To understand why the protected in-gap states in the ES of the cut $A^{m_{2}}_{m_{1}}$ are in the range of $\left[1/m_{1},1-1/m_{1}\right]$, we need the following lemma. Given two Hermitian matrices
$A$ and $B$, with eigenvalues $a_i$'s and $b_i$'s, consider their sum $M=A+B$, for each $a_{i}$, there must be an eigenvalue of $M$ which satisfies $a_{i}+\min\{b_{i}\}\leqslant m^{\star}\leqslant a_{i}+\max\{b_{i}\}$. With this, let us discuss the eigenvalues of each block of the correlation operator shown in Eq.\eqref{eq:correlator}, $i.e.$, the eigenvalues of $\frac{1}{m_{1}}\sum_{l=0}^{m_{1}-1}D_{l}^{r}$. If we set $A=D_{0}^{r}+D_{1}^{r}$ and $B=\sum_{l=2}^{m_{1}-1}D_{l}^{r}$, there will be at least $|\mathrm{dim}(D^{r}_{0})-\mathrm{dim}(D^{r}_{1})|$ eigenstates of $A$ with eigenvalue 1. Since $\min\{b_{i}\}=0$ and $\max\{b_{i}\}=m_{1}-2$, there are at least $|\mathrm{dim}(D^{r}_{0})-\mathrm{dim}(D^{r}_{1})|$ eigenstates of $\frac{1}{m_{1}}\sum_{l=0}^{m_{1}-1}D_{l}^{r}$ with eigenvalues in the range of $\left[(0+1)/m_{1},(m_{1}-2+1)/m_{1}\right]$, which is just $\left[1/m_{1},1-1/m_{1}\right]$. Since we can set $A$ to be the sum of any two $D^{r}_{l}$'s, there are at least $\max_{l,l^{\prime}}|\mathrm{dim}(D^{r}_{l})-\mathrm{dim}(D^{r}_{l^{\prime}})|$ protected in-gap states in the range of $\left[1/m_{1},1-1/m_{1}\right]$. Appendix C of Ref \onlinecite{fang2013entanglement} relates $\mathrm{dim}(D^{r}_{l})$ to the $\mathbb{Z}^{n}$ index, but the proof is technical and will not be reviewed here.
\begin{table}[ht]
\centering
\caption{Number of protected in-gap states in the entanglement spectrum for some symmetric cuts $A^{m_{2}}_{1/m_{1}}$ in $C_{n}$-symmetric insulators used in the main text in terms of the $\mathbb{Z}^n$ index. The table is partially adapted from Ref \onlinecite{fang2013entanglement}.} 
\label{tab:tab1ref8}
\renewcommand{\arraystretch}{1.5}
\begin{tabular}{p{0.12\columnwidth}p{0.12\columnwidth}p{0.12\columnwidth}p{0.64\columnwidth}}
\hline\hline
$n$ & $m_{1}$ & $m_{2}$ & Number of protected in-gap states\\
\hline
2 & 2 & 1 & $|z_{0}-z_{1}|$\\
3 & 3 & 1 & $\max _{i, j=0,1,2}\left|z_{i}-z_{j}\right|$\\
4 & 2 & 2 &  $\left|z_{0}-z_{2}\right|+\left|z_{1}-z_{3}\right|$\\
4 & 4 & 1 & $\max _{i, j=0,1,2,3}\left|z_{i}-z_{j}\right|$\\
\hline\hline
\end{tabular}
\end{table}
\section{\label{app:ingap}Corner-induced filling anomaly in general cases}

\subsection{\texorpdfstring{$C_2$}{C2} symmetric TCIs}
In $C_{2}$ symmetric TCIs, let us consider the ES for the $A_{1/2}^1$ cut in an infinite rectangular lattice with the $C_{2}$ rotation center localized at Wyckoff position $b$. According to Ref. \onlinecite{fang2013entanglement}, the number of protected in-gap states with eigenvalue $1/2$ in the ES is
\begin{equation}
\mathrm{no}.\frac{1}{2}_{p,2,1}^{(2)} = |z_0 - z_1| = |n_b^0 - n_b^1|.
\end{equation}
Since $n_b = n_b^0 + n_b^1$ has the same parity as $|n_b^0 - n_b^1|$, we have $n_{b}\ \mathrm{mod} \ 2=\mathrm{no}.\frac{1}{2}_{p,2,1}^{(2)}\ \mathrm{mod} \ 2$. Thus, the corner-induced filling anomaly in Eq.\eqref{eq:c236fa} can be written as
\begin{equation}
\label{app:C2forC6}
\eta^{(2)} =- n_{b} \ \mathrm{mod} \ 2=n_{b} \ \mathrm{mod} \ 2=\mathrm{no}.\frac{1}{2}_{p,2,1}^{(2)} \ \rm{mod} \ 2.
\end{equation}

\subsection{\texorpdfstring{$C_3$}{C3} symmetric TCIs}
In $C_3$ symmetric TCIs, let us consider the $A^{1}_{1/3}$ cut. As discussed in Ref. \onlinecite{fang2013entanglement}, we can have $\rm{max}_{i,j=1,2,3}|z_{i}-z_{j}|$ protected in-gap states in the range of $\left[1/3,2/3\right]$. If we denote $\rm{max}_{i,j=1,2,3} {z_{i}}$ as $z^{first}$, $\rm{min}_{i,j=1,2,3} {z_{i}}$ as $z^{third}$ and the second biggest of $z_1, z_2, z_3$ as $z^{second}$, then it is easy to see $\rm{max}_{i,j=1,2,3}|z_{i}-z_{j}|=z^{first}-z^{third}$. Furthermore, we can determine the number of protected states with eigenvalue $1/3$ from the reduced correlation function for the cut $A_{1/3}^1$
\begin{equation}
C(A)=\frac{1}{3}(D_{0}+D_{1}+D_{2})
\end{equation}
where each $D_i^{2}=D_{i}$ is a projector of the subspace in which states pick up a phase $e^{i(2\pi l+\pi F)/3}$ under $C_{3}$ rotation.
Since we have $m_{2}=1$ for the cut $A^{1}_{1/3}$, the superscript $p$ in Eq. \eqref{eq:correlator} can only be 0, and thus has been suppressed in the above equation. If one of the projectors has dimension greater than all the others, there will be protected in-gap states with eigenvalue $1/3$, of which the number is
\begin{equation}
\begin{aligned}
\mathrm{no}.\frac{1}{3}_{p,3,1}^{(3)}=\mathrm{dim}(D^{first})-\mathrm{dim}(D^{second})=z^{first}-z^{second}.
\end{aligned}
\end{equation}

If we choose the $C_{3}$ rotation center to be at Wyckoff position $b$, then we have,
\begin{equation}
\begin{array}{l}
\mathrm{no}.\frac{1}{3}_{p,3,1}^{(3)}=z^{first}-z^{second}=n^{first}_{b}-n^{second}_{b},
\\
\\
\mathrm{no}.\left[1/3,2/3\right]_{p,3,1}^{(3)}=z^{first}-z^{3nd}=n^{first}_{b}-n^{third}_{b},
\end{array}
\end{equation}
where $n^{first}_{b}$, $n^{second}_{b}$, and $n^{third}_{b}$ are the maximum, middle, and minimum of $\{n_{b}^{0},n_{b}^{1},n_{b}^{2}\}$, respectively. Then, we have
\begin{equation}
\label{eq:A6}
\begin{aligned}
n_{b}&=n^{first}_{b}+n^{second}_{b}+n^{third}_{b}
\\
&=3n^{first}-(n^{first}_{b}-n^{second}_{b})-(n^{first}_{b}-n^{third}_{b})
\\
&=3n^{first}-\mathrm{no}.\frac{1}{3}_{p,3,1}^{(3)}-\mathrm{no}.\left[1/3,2/3\right]_{p,3,1}^{(3)}.
\end{aligned}
\end{equation}
From Eq.\eqref{eq:A6} and Eq. \eqref{eq:c236fa}, we derive the corner-induced filling anomaly as
\begin{equation}
\label{app:C3forC6}
\begin{aligned}
\eta^{(3)}&=-n_{b}\ \rm{mod} \ 3
\\
&=\left(\mathrm{no}.\frac{1}{3}_{p,3,1}^{(3)}+\mathrm{no}.\left[1/3,2/3\right]_{p,3,1}^{(3)}\right) \ \rm{mod} \ 3.
\end{aligned}
\end{equation}

\subsection{\texorpdfstring{$C_6$}{C6} symmetric TCIs} 
In $C_{6}$ symmetric TCIs, since the corner-induced filling anomaly depends on $n_{b}$ and $n_{c}$, in order to write the corner-induced filling anomaly in terms of the number of protected in-gap states, we need to study the ES for the cut $A_{1/3}^{1}$ in an infinite lattice with $C_{3}$ rotation center localized at Wyckoff position $b$, and the ES for the cut $A_{1/2}^{1}$ in the same lattice with $C_{2}$ rotation center localized at Wyckoff position $c$. Since we have already discussed the ES for these two cuts in the $C_{2}$ and $C_{3}$ symmetric cases, we can substitute Eq. \eqref{app:C3forC6} and Eq. \eqref{app:C2forC6} into Eq. \eqref{eq:c236fa} and write the corner-induced filling anomaly as
\begin{equation}
\begin{aligned}
\eta^{(6)} = &-2\left(\mathrm{no}.\frac{1}{3}_{p,3,1}^{(6)}+\mathrm{no}.\left[1/3,2/3\right]_{p,3,1}^{(6)} \right) 
\\
&+3\mathrm{no}.\frac{1}{2}_{p,2,1}^{(6)}\ \rm{mod} \ 6.
\end{aligned}
\end{equation}

\section{\label{app:parity} The parity of $\frac{1}{4}\left(\mathrm{no}.\frac{1}{2}\right)_{corner}$ in the cut $A^{1}_{1/4}$ for the $C_{4}$ symmetric case}
First, we point out that the term containing $n_{b,d}^{\prime}$ in Eq. \eqref{eq: onehalf0} does not affect the parity of $\frac{1}{4}\mathrm{no}.\frac{1}{2}$. What is more, $n_{b,s}+n_{b,t}$ and $n_{c,s}$ are either both even or both odd. This is because $n_{b}=n_{b,s}+2n_{b,d}+3n_{b,t}+4n_{b,q}$, and $n_{c}=n_{c,s}+2n_{c,d}$ are either both even or both odd due to the zero bulk polarization requirement. Since $n_{b,s}+n_{b,t}=\frac{1}{4}(\mathrm{no}.\frac{1}{4}+\mathrm{no}.\frac{3}{4})$, we can see that when $\frac{1}{4}(\mathrm{no}.\frac{1}{4}+\mathrm{no}.\frac{3}{4})$ is even, the parity of $\frac{1}{4}\left(\mathrm{no}.\frac{1}{2}\right)_{corner}$ is the same as the parity of $\frac{1}{4}\mathrm{no}.\frac{1}{2}$, and when $\frac{1}{4}(\mathrm{no}.\frac{1}{4}+\mathrm{no}.\frac{3}{4})$ is odd, the parity of $\frac{1}{4}\left(\mathrm{no}.\frac{1}{2}\right)_{corner}$ is opposite to the parity of $\frac{1}{4}\mathrm{no}.\frac{1}{2}$. Then, we can conclude that the parity of $\frac{1}{4}\left(\mathrm{no}.\frac{1}{2}\right)_{corner}$ is always equal to the parity of $\frac{1}{4}(\mathrm{no}.\frac{1}{2}-\mathrm{no}.\frac{1}{4}-\mathrm{no}.\frac{3}{4})$.
\section{\label{app:CnDeriv} Corner-induced filling anomaly in the Zero Length Limit}
\subsection{\texorpdfstring{$C_2$}{C2} symmetric TCIs}
\begin{figure}[h]
\includegraphics[width=1\columnwidth]{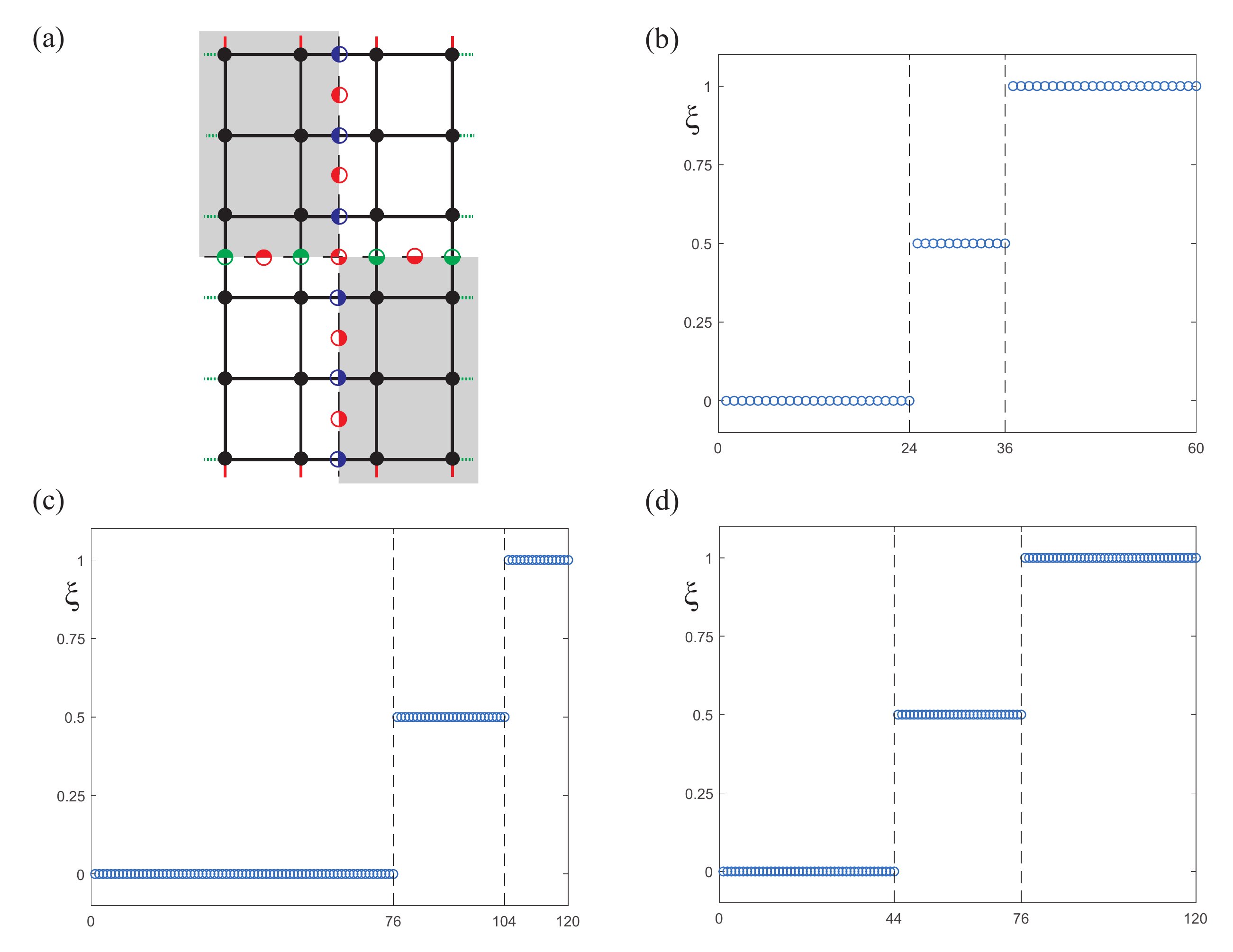}
\caption{\label{fig:c2latticecut} (a) Schematic illustration of the corner-crossing cut in a periodic lattice. The red solid (green dotted) lines on edges indicate the hopping between top and bottom (right and left) edges. Wannier orbitals at $b$, $c$, and $d$ are represented by red, blue, and green dots respectively. ES of the cross corner cut shown in Fig. \ref{fig:c2latticecut}(a) in 6 x 10 lattice with periodic boundary conditions for the (b) $h_{1d}^{(2)}$; (c) $h_{1b}^{(4)}$; (d) $h_{2c}^{(4)}$ in Ref. \onlinecite{benalcazar2018quantization}. }
\end{figure}

In $C_{2}$ symmetric lattices, we study the cut shown in Fig. \ref{fig:c2latticecut}(a).
In $C_2$ symmetric TCIs, we can have only singlets and doublets of Wannier orbitals. On the edges of the cut, the singlets of Wannier orbitals localized at $b$, $c,$ and $d$ contribute in-gap states with eigenvalue $\frac{1}{2}$ to the ES, while the doublets do not contribute in-gap states. At the corner-crossing of this cut, only the singlets of Wannier orbitals localized at $b$ contribute in-gap states with eigenvalue $\frac{1}{2}$. As such, the number of in-gap states with eigenvalue $\frac{1}{2}$ in the entanglement spectrum is
\begin{equation}
\mathrm{no}.\frac{1}{2} = (2L_y+2L_x - 4)n_{b,s} + 2L_y n_{c,s} + 2L_x n_{d,s}
\label{eq:c2num}
\end{equation}
where $L_x$ and $L_y$ are the number of lattice points on the horizontal and vertical edges respectively. 

Expressing Eq. \eqref{eq:c2num} in a more useful way, we see
\begin{equation}
     \frac{1}{4}(\mathrm{no}.\frac{1}{2}^{(2)}) = \frac{L_x}{2}(n_{b,s} + n_{d,s}) + \frac{L_y}{2}(n_{b,s} + n_{c,s}) - n_{b,s}.
    \label{eq:c2num2}
\end{equation}
We can see that the left-hand side of Eq. \eqref{eq:c2num2} must have the same parity as $n_b=n_{b,s}+2n_{b,d}$, because the first two terms on the left-hand side are even if we choose the corner-crossing to be located at Wyckoff position $b$ and require the lattice to be $C_{2}$ symmetric. Thus, we can determine the corner-induced filling anomaly by looking at the number of in-gap states with eigenvalue $\frac{1}{2}$ in the corner-crossing cut:
\begin{equation}
\label{app: c2fa0}
\eta^{(2)} = n_{b}\ \rm{mod}\ 2 =\frac{1}{4}(\mathrm{no}.\frac{1}{2}^{(2)})\ \rm{mod}\ 2.
\end{equation}

These results are numerically confirmed through the ES shown in Fig \ref{fig:c2latticecut}. In Fig \ref{fig:c2latticecut}(b), there are 12 eigenstates with eigenvalue $\frac{1}{2}$. By substituting $\mathrm{no}.\frac{1}{2}=12$ into Eq. \eqref{app: c2fa0}, we can determine the corner-induced filling anomaly is $\eta^{(2)} =1$, which matches the corner-induced filling anomaly of $h^{(2)}_{1d}$. In Fig \ref{fig:c2latticecut}(c) there are 28 in-gap states with eigenvalue $\frac{1}{2}$, and in Fig \ref{fig:c2latticecut} (d) there are 32 in-gap states with eigenvalue $\frac{1}{2}$, which leads to a corner-induced filling anomaly of 1 and 0, respectively. This matches the the corner-induced filling anomaly of $h^{(4)}_{1b}$ and $h^{(4)}_{2c}$.

\subsection{\texorpdfstring{$C_3$}{C3} symmetric TCIs}
\begin{figure}[h]
\includegraphics[width=1 \columnwidth]{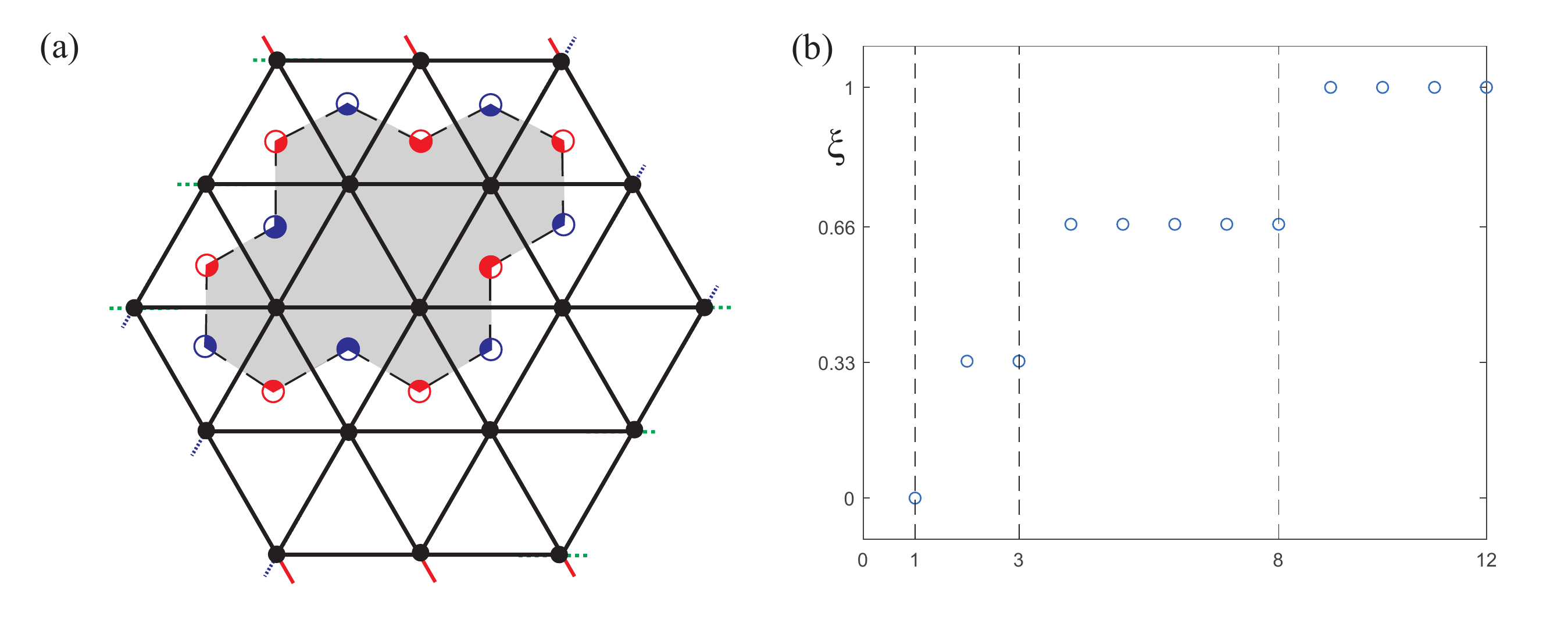}
\caption{\label{fig:C3} (a) Schematic illustration of the  cut we use in $C_6$ symmetric TCIs in a periodic lattice. The red solid (green dotted, blue dashed) lines on edges indicate the hopping between edges. Wannier orbitals at $b$ and $c$ are represented by red and blue dots respectively. ES of the cut shown in Fig. \ref{fig:C3}(a) for (b) $h_{2b}^{(3)}$ ($h_{2c}^{(3)}$ has the same ES) in Ref. \onlinecite{benalcazar2018quantization}}
\end{figure}

For $C_3$ symmetric lattices, we study the cut shown in Fig. \ref{fig:C3}(a). 
Here, only singlets and doublets of $b$ and $c$ contribute in-gap states with eigenvalues $1/3$ and $2/3$ to the ES. Following the method discussed previously, we can have two equations for this cut:
\begin{equation}
\label{eq:c3num}
\begin{array}{l}
\begin{aligned}
\mathrm{no}.\frac{1}{3}^{(3)}=\ & 2(L_{cut}-1)(n_{b,d}+n_{c,d})
\\ & + (2L_{cut}+1)(n_{b,s}+n_{c,s}),
\end{aligned} \\
\begin{aligned}
\mathrm{no}.\frac{2}{3}^{(3)}=\ & 2(L_{cut}-1)(n_{b,s}+n_{c,s})
\\&+(2L_{cut}+1)(n_{b,d}+n_{c,d}).
\end{aligned}
\end{array}
\end{equation} Substituting Eq.\eqref{eq:c3num} into Eq.\eqref{eq:c236fa}, we find that the corner-induced filling anomaly can be expressed from the entanglement spectrum as:
\begin{equation}
\label{eq:c3fa}
\eta^{(3)}= \frac{1}{3}(\mathrm{no}.\frac{1}{3}^{(3)}-\mathrm{no}.\frac{2}{3}^{(3)}) \ \rm{mod} \ 3.
\end{equation}

The numerical results are shown in Fig. \ref{fig:C3} for the two models in Ref. \onlinecite{benalcazar2018quantization}. Note that both of these models have the same ES. There are two in-gap states with eigenvalue $\frac{1}{3}$ and five in-gap states with eigenvalue $\frac{2}{3}$, which yields a corner-induced filling anomaly of $\eta^{(3)} =1$, as we expect.

\subsection{\texorpdfstring{$C_6$}{C6} symmetric TCIs} 
\begin{figure}[h]
\includegraphics[width=1\columnwidth]{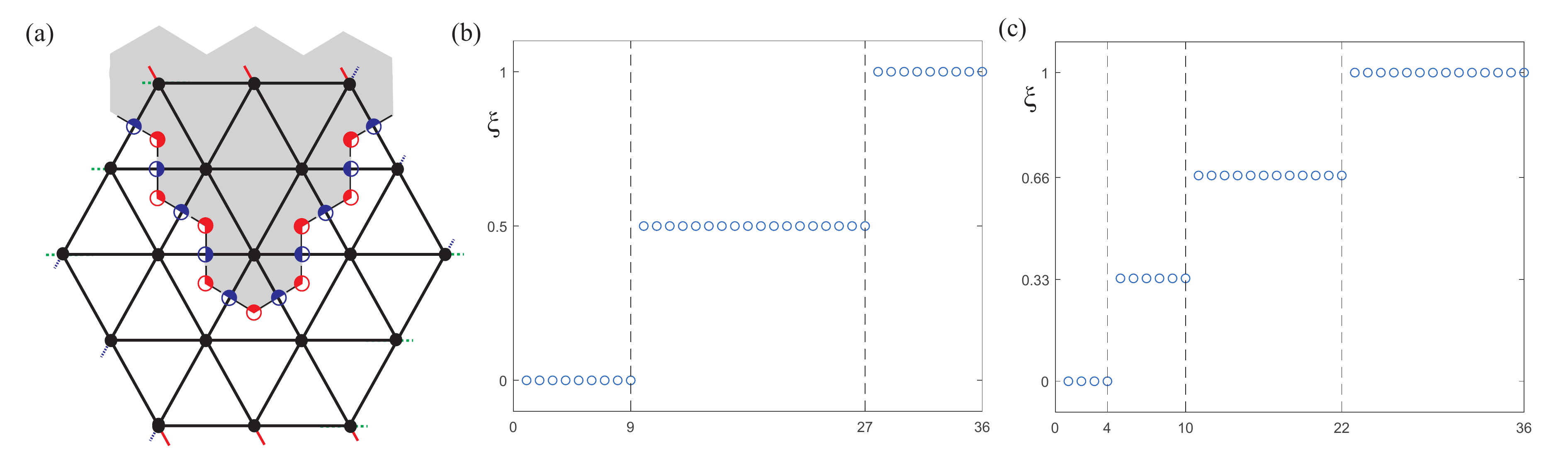}
\caption{\label{fig:C6} (a) Schematic illustration of the cut we use in $C_6$ symmetric TCIs in the periodic lattice. The red solid (green dotted) lines on edges indicate the hopping between top and bottom (right and left) edges. Wannier orbitals at $b$ and $c$ are represented by red and blue  dots respectively. ES of cut shown in (a) for the (b) $h_{3c}^{(6)}$; (c) $h_{4b}^{(6)}$ in Ref. \onlinecite{benalcazar2018quantization}}
\end{figure}

For $C_6$ symmetric TCIs, we study the cut shown in Fig. \ref{fig:C6}(a). 
Along this cut, singlets and doublets of Wannier orbitals localized at Wyckoff position $b$ contribute in-gap states with eigenvalues $1/3$ and $2/3$, and singlets of Wannier orbitals localized at Wyckoff position $c$ contribute in-gap states with eigenvalue $\frac{1}{2}$. We then have three equations:
\begin{equation}
\label{eq:c6num}
\begin{array}{l}
\mathrm{no}.\frac{1}{3}^{(6)}=3(L_{cut}-1)n_{b,d}+3(L_{cut}+1)n_{b,s},
\\
\\
\mathrm{no}.\frac{2}{3}^{(6)}=3(L_{cut}-1)n_{b,s}+3(L_{cut}+1)n_{b,d},
\\
\\
\mathrm{no}.\frac{1}{2}^{(6)}=6L_{cut}n_{c,s},
\end{array}
\end{equation}
where $L_{cut}$ represents the number of lattice points along one edge of the cut.

Representing Eq. \eqref{eq:c236fa} in another way, we find 
\begin{equation}
\eta^{(6)} = (2n_{b,s} - 2n_{b,d} + 3n_{c,s}) \ \rm{mod} \ 6.
\label{app:C6FA1}
\end{equation}
Substituting Eq. \eqref{eq:c6num} into Eq. \eqref{app:C6FA1}, we find that as long as $L_{cut}$ is odd, we can write the corner-induced filling anomaly as
\begin{equation}
\label{eq: c6fa2}
\eta^{(6)}= \frac{1}{6}(2\mathrm{no}.\frac{1}{3}-2\mathrm{no}.\frac{2}{3}+3\mathrm{no}.\frac{1}{2}) \ \rm{mod} \ 6.
\end{equation}
We confirm this result numerically for two TCIs shown in Fig. \ref{fig:C6}. In Fig. \ref{fig:C6}(b) there are 18 in-gap states with eigenvalue $\frac{1}{2}$. By substituting this into Eq. \eqref{eq: c6fa2}, we find $\eta^{(6)} = 3$, which matches the filling anomaly of $h_{3c}^{(6)}$.  In Fig. \ref{fig:C6}(c) there are 6 in-gap states with eigenvalue $\frac{1}{3},$ and 12 in-gap states with eigenvalue $\frac{2}{3}$. Again, by substituting this into Eq. \eqref{eq: c6fa2}, we get $\eta^{(6)} = 4$, which matches the filling anomaly of $h_{4b}^{(6)}$.

\end{document}